\documentclass[12pt,a4paper]{article}
\usepackage[T1]{fontenc}
\usepackage{lmodern}
\usepackage[margin=23mm]{geometry}
\usepackage{amsmath,amssymb,bm,graphicx,booktabs,array,tabularx}
\usepackage{authblk,microtype,placeins,float}
\usepackage[font=small,labelfont=bf,labelsep=period]{caption}
\usepackage[numbers,sort&compress]{natbib}
\usepackage[colorlinks=true,allcolors=blue]{hyperref}
\numberwithin{figure}{section}
\numberwithin{table}{section}

\newsavebox{\paperTableBox}
\newcommand{\fittable}[1]{%
  \sbox{\paperTableBox}{#1}%
  \ifdim\wd\paperTableBox>\linewidth
    \resizebox{\linewidth}{!}{\usebox{\paperTableBox}}%
  \else
    \usebox{\paperTableBox}%
  \fi
}
\title{Free-boundary multi-fluid tokamak equilibria\\with toroidal rotation and species separation}
\hypersetup{pdftitle={Free-boundary multi-fluid tokamak equilibria with toroidal rotation and species separation}}
\author[1,2]{Xingyu Li}
\author[2\textdagger]{Huasheng Xie}
\author[1]{Lai Wei}
\author[1\textdagger]{Zheng-Xiong Wang}
\affil[1]{\small Key Laboratory of Materials Modification by Laser, Ion, and Electron Beams of the Ministry of Education, School of Physics, Dalian University of Technology, Dalian 116024, China}
\affil[2]{\small Beijing VeloAlpha Technology Co., Ltd., Beijing, 100080, China}
\date{}
\begin{document}
\raggedbottom
\setlength{\emergencystretch}{1.5em}
\maketitle
\begin{NoHyper}
{\renewcommand{\thefootnote}{\textdagger}
\footnotetext{Corresponding authors. E-mail: huashengxie@gmail.com (H. Xie), zxwang@dlut.edu.cn (Z.-X. Wang).}}
\end{NoHyper}
\vspace{-\baselineskip}

\begin{abstract}
We develop MFEQ-FB, an axisymmetric solver that calculates species separation and the plasma boundary self-consistently by extending a reduced rotating multi-fluid model to free-boundary equilibria. Numerical verification recovers static magnetohydrodynamic (MHD) equilibria for nonrotating multi-ion plasmas and rotating MHD equilibria for single-ion plasmas. In an ITER-like pure deuterium--tritium (DT) case at an on-axis angular velocity of \(70\ \mathrm{krad/s}\), the multi-fluid and rotating MHD boundaries nearly coincide, while the outboard T/D density ratio exceeds the inboard value by \(4.8\%\) on the magnetic surface with normalised poloidal flux \(x=0.50\) at \(Z=0.50\ \mathrm m\). At \(15\ \mathrm{MA}\) and \(252\ \mathrm{krad/s}\) on axis, separate scans vary the ion composition and ion-to-electron temperature ratio while holding the coil currents, angular-velocity profile and reference profiles of electron density, ion mass density and total thermal pressure fixed. These inputs give a common rotating MHD equilibrium, whereas the multi-fluid equilibria exhibit different species distributions and boundaries. In pure DT, reducing the ion-to-electron temperature-ratio profile, with its on-axis value decreasing from \(0.888\) to \(0.370\), increases the root-mean-square distance between the two models' last closed flux surfaces from \(36\) to \(84\ \mathrm{mm}\). Thus, boundary agreement can coexist with D--T separation on the same magnetic surface, and resolving species-dependent thermal pressures reveals free-boundary responses that are absent from the common rotating MHD reference equilibrium.
\end{abstract}

\section{Introduction}
\label{sec:1}

Toroidal rotation changes the pressure distribution on magnetic surfaces and the magnetic geometry of tokamak plasmas. Toroidal rotation arises intrinsically and is also observed in plasmas heated by neutral beams or ion-cyclotron waves~\cite{rice1998centralrotation,field2011rotation,lin2008modeconversion,bush1997tftr}. Predictions for ITER depend on the heating conditions and the momentum-transport model~\cite{budny2009iter,budny2012alpha,paul2017rotation,chrystal2020predicting}. When the toroidal speed is comparable to the ion thermal speed, centrifugal forces redistribute the plasma along magnetic surfaces. In a mixture of ion species, the density response also depends on species mass, charge and temperature, coupling fuel composition to the equilibrium pressure.

Rotating magnetohydrodynamic (MHD) theory describes the centrifugal modification of the total pressure and magnetic equilibrium~\cite{hameiri1983equilibrium,maschke1980exact}. Flow has been incorporated into both fixed-boundary and free-boundary MHD solvers~\cite{guazzotto2004flow,li2021spectralrotation,chen2022freeboundaryflow,ftorijadaza2024flipec}. VEQ-R uses a compact spectral representation to resolve rotation-induced changes in the internal magnetic geometry of fixed-boundary MHD equilibria~\cite{li2026veqr}. Resolving the individual species introduces an additional question: how does their redistribution relate to the change in magnetic geometry?

Rotation-induced impurity-density variation has been studied in tokamaks~\cite{wesson1997poloidal}. Multi-fluid theories describe species flows and their coupling to the fields~\cite{steinhauer1999formalism,thyagaraja2006flows}, with numerical implementations including FLOW2 and axisymmetric two-fluid and three-fluid models~\cite{guazzotto2015flow2,ishida2010twofluid,ishida2015threefluid}. Galeotti et al.~\cite{galeotti2011multiion} formulated rotating multi-ion equilibria using species-dependent surface functions and implemented far-field boundary conditions that account for external coils. Xie et al.~\cite{xie2026reduced} developed a reduced multi-fluid model with profiles labelled by poloidal flux, retaining toroidal ion inertia and electrostatic polarisation, and applied it to proton--boron equilibria with prescribed boundaries. This model was implemented in the spectral solver VEQ-MF and used in parameter scans of fixed-boundary p--$^{11}$B equilibria to study species redistribution, fusion power and bremsstrahlung losses~\cite{li2026veqmf}.

In a free-boundary equilibrium, the plasma current and external coil currents determine the poloidal flux and plasma boundary~\cite{lackner1976equilibria}. Free-boundary methods have been applied to shaped plasmas and divertor configurations~\cite{hofmann1988fbt,jeon2015tes,heumann2015cedres,farmakalides2025cratos}, and calculated MAST-U equilibria have been compared with experimental reconstructions~\cite{pentland2025validation}. Species redistribution enters this problem through the total thermal pressure and its flux derivative in the toroidal-current source. Even at fixed total plasma current, a change in the spatial current distribution can change the magnetic surfaces. The species distributions and boundary must therefore be determined together to evaluate this response.

We extend this reduced multi-fluid (MF) model to free-boundary equilibria and develop MFEQ-FB, an axisymmetric solver that determines species densities, electrostatic potential and the plasma boundary self-consistently for prescribed external coil currents and total plasma current. The ITER-like deuterium--tritium (DT) and DT--He applications examine D--T separation when the MF and rotating MHD boundaries nearly coincide, and the effects of composition and ion-to-electron temperature ratio when the rotating MHD reference is unchanged. We compare separation estimates on rotating MHD surfaces with self-consistent MF values and compare the outboard boundary positions predicted by the two models.

Sections~\ref{sec:2} and \ref{sec:3} present the model, numerical method and verification. Section~\ref{sec:4} examines rotation, ion composition and temperature partition. Section~\ref{sec:5} discusses their implications for species separation and free-boundary equilibrium.

\FloatBarrier
\section{Equilibrium model and numerical method}
\label{sec:2}

\FloatBarrier
\subsection{Rotating multi-fluid model}
\label{sec:2-1}

We consider a steady, axisymmetric plasma consisting of electrons and multiple ion species in cylindrical coordinates \((R,\phi,Z)\). The reduced multi-fluid model of Xie et al. retains toroidal ion inertia and the electrostatic polarisation potential, while neglecting electron inertia and poloidal-flow inertia. Each species has an isotropic pressure, and its temperature is constant on a magnetic surface. Under these assumptions, the species-density relations are coupled to a generalised Grad-Shafranov equation \cite{xie2026reduced}.

The prescribed reference-density and temperature profiles \(N_s(x)\) and \(T_s(x)\) and the common ion angular-velocity profile \(\Omega_i(x)\) are labelled by poloidal magnetic flux. This approximation neglects the finite-orbit-width correction \(Y_s-\psi\) to the drift-surface label of each species. The normalised poloidal flux is

\[
x=\frac{\psi-\psi_a}{\psi_b-\psi_a},\qquad 0\leq x\leq 1,
\tag{2.1}
\label{eq:2-1}
\]

where \(\psi_a\) and \(\psi_b\) are the poloidal fluxes at the magnetic axis and the last closed flux surface (LCFS), respectively. The reference radius \(R_0\) is prescribed, and the equilibrium solution determines the magnetic-axis position. The mass and charge of species \(s\) are \(m_s\) and \(Z_se\), respectively, with \(e>0\) and \(Z_e=-1\). Temperatures supplied in electronvolts are converted to thermal energies through \(\theta_s=eT_{s,\mathrm{eV}}\).

Appendix~\ref{app:ion-label} assesses \(Y_s-\psi\) and the resulting first-order changes in the reference profiles for the ten \(15\ \mathrm{MA}\) application cases.

At a given \((R,x)\), the species density is

\[
\begin{aligned}
n_s(R,x)
&=N_s(x)\exp\!\left[
\frac{W_s(R,x)-Z_se\Phi(R,x)}{\theta_s(x)}
\right],\\
W_s(R,x)
&=\frac{1}{2}m_s\Omega_i^2(x)(R^2-R_0^2),\quad s\ne e,
\qquad W_e=0,
\end{aligned}
\tag{2.2}
\label{eq:2-2}
\]

where \(\Phi\) is the electrostatic polarisation potential. Quasineutrality determines \(\Phi\), and summing the species thermal pressures gives the total thermal pressure

\[
\sum_s Z_sn_s(R,x)=0,\qquad
P_{\mathrm{MF}}(R,x)=\sum_s n_s(R,x)\theta_s(x).
\tag{2.3}
\label{eq:2-3}
\]

The reference densities satisfy \(N_e(x)=\sum_{\mathrm{ions}}Z_sN_s(x)\). Formally setting \(R=R_0\) gives \(\Phi(R_0,x)=0\) and \(n_s(R_0,x)=N_s(x)\), which normalises the density relation at every \(x\), including on inner surfaces with no intersection with the cylinder \(R=R_0\). Quasineutrality couples the species densities through the common potential. The pressure therefore depends on both \(R\) and \(x\).

For a given poloidal-flux distribution, the quasineutrality equation determines \(\Phi\), from which the species densities, pressure and pressure derivative are evaluated. The pressure derivative enters the Grad-Shafranov current source, so these quantities must be solved self-consistently with the poloidal flux.

The MHD-limit tests and the \(10\ \mathrm{MA}\) grid-refinement study in Section~\ref{sec:3} use prescribed reference-density, temperature and common ion angular-velocity profiles of the form

\[
Q(x)=Q_{\mathrm{edge}}+
\left(Q_{\mathrm{axis}}-Q_{\mathrm{edge}}\right)\left(1-x^{\beta_Q}\right)^{\alpha_Q},
\tag{2.4}
\label{eq:2-4}
\]

where \(Q\in\{N_s,T_s,\Omega_i\}\), \(Q_{\mathrm{axis}}\) and \(Q_{\mathrm{edge}}\) are the values at \(x=0\) and \(x=1\), and \(\alpha_Q\) and \(\beta_Q\) control the profile shape. Equal axis and edge values give a constant profile. The separate \(15\ \mathrm{MA}\) pure-DT benchmark in Section~\ref{sec:3-2} uses the finite-edge-density profiles specified in Section~\ref{sec:4-3}.

The Grad-Shafranov source contains \(-FF'\), with \(F=RB_\phi\) and baseline profile
\[
F_b(x;\gamma)=F_{\mathrm{vac}}\left[1-\gamma\left(1-x^{\beta_F}\right)^{\alpha_F}\right].
\tag{2.5}
\label{eq:2-5}
\]

Here, \(F_{\mathrm{vac}}\) is the vacuum value of the toroidal-field function, and \(\alpha_F\) and \(\beta_F\) set the shape of \(F_b\). When the baseline toroidal-field profile is used directly, \(F=F_b\), as in the \(15\ \mathrm{MA}\) applications. Section~\ref{sec:4-1} gives the rescaling of \(F_b^2-F_{\mathrm{vac}}^2\) used in the \(11\ \mathrm{MA}\) configuration. Section~\ref{sec:2-2} describes calculations with prescribed total current or prescribed \(\gamma\).

The rotating MHD model uses the total reference thermal pressure \(P_0\) and reference ion mass density \(\rho_0\) calculated from the species profiles

\[
P_0(x)=\sum_sN_s(x)\theta_s(x),\qquad
\rho_0(x)=\sum_{\mathrm{ions}}m_sN_s(x).
\tag{2.6}
\label{eq:2-6}
\]

The derivative with respect to the normalised poloidal flux follows from the sum in Eq.~\eqref{eq:2-6}

\[
P_{0,x}(x)=\sum_s\left(N_{s,x}\theta_s+N_s\theta_{s,x}\right).
\]

All ions share the prescribed angular velocity, \(\Omega_s(x)=\Omega_i(x)=\Omega(x)\). For the rotating-MHD reference, the ratio of thermal pressure to mass density is constant on each magnetic surface and equals \(P_0(x)/\rho_0(x)\). Force balance along the surface then gives

\[
P_{\mathrm{MHD}}(R,x)=P_0(x)\exp\!\left[
\frac{\rho_0(x)\Omega^2(x)(R^2-R_0^2)}{2P_0(x)}
\right].
\tag{2.7}
\label{eq:2-7}
\]

Each pressure expression is evaluated using the poloidal flux of its own equilibrium solution.

When the reference densities share a factor \(h(x)\) that vanishes at \(x=1\), as in the MHD-limit tests and the \(10\ \mathrm{MA}\) grid-refinement study in Section~\ref{sec:3}, we cancel this factor in the quasineutrality equation and in the ratio \(\rho_0/P_0\) in the rotating MHD exponent. The factored expressions at \(x=1\) take their continuous limits from inside the plasma. The physical densities and pressures retain \(h\).

The static- and rotating-MHD limits are presented in Section~\ref{sec:3-1}.

The local and reference thermal Mach numbers for ion species \(s\) are

\[
M_s(R,x)=\frac{R|\Omega_s(x)|}{\sqrt{\theta_s(x)/m_s}},
\qquad
M_{s,\mathrm{ref}}=\frac{R_0|\Omega_0|}{\sqrt{\theta_s(0)/m_s}}.
\tag{2.8}
\label{eq:2-8}
\]

Here \(\Omega_0=\Omega_s(0)\), and \(R_0=6.20\ \mathrm m\) in Section~\ref{sec:4}. The toroidal speed at the computed magnetic axis is \(v_{\mathrm{ax}}=R_{\mathrm{ax}}|\Omega_0|\), where \(R_{\mathrm{ax}}\) is the magnetic-axis radius in that equilibrium. The on-axis Mach number uses this speed and \(\theta_s(0)\).

\FloatBarrier
\subsection{Free-boundary equilibrium and plasma-current constraint}
\label{sec:2-2}

The magnetic field is written as

\[
\mathbf B=\nabla\psi\times\nabla\phi+F(\psi)\nabla\phi,
\tag{2.9}
\label{eq:2-9}
\]

where \(\psi\) is the poloidal magnetic flux per radian and \(F\) is the toroidal-field function defined in Section 2.1. In this convention the generalised Grad-Shafranov equation is

\[
\Delta^*\psi
=-\mu_0R^2\left(\frac{\partial P}{\partial\psi}\right)_R-FF',
\qquad
\Delta^*=R\frac{\partial}{\partial R}
\left(\frac{1}{R}\frac{\partial}{\partial R}\right)
+\frac{\partial^2}{\partial Z^2}.
\tag{2.10}
\label{eq:2-10}
\]

Here \(\mu_0\) is the vacuum permeability, \(F'=\mathrm dF/\mathrm d\psi\), and \(P\) is evaluated from the multi-fluid or rotating-MHD pressure expression, with both models using the same macroscopic plasma inputs. The toroidal current density separates into pressure-derivative and toroidal-field-function contributions.

\[
\begin{aligned}
J_\phi&=J_P+J_F, &
J_P&=R\left(\frac{\partial P}{\partial\psi}\right)_R, &
J_F&=\frac{FF'}{\mu_0R},\\
\left(\frac{\partial P}{\partial\psi}\right)_R
&=\frac{P_x}{\psi_b-\psi_a},&
P_x&=\left(\frac{\partial P}{\partial x}\right)_R.
\end{aligned}
\tag{2.11}
\label{eq:2-11}
\]

At fixed \(R\), we implicitly differentiate quasineutrality to obtain \(\Phi_x=(\partial\Phi/\partial x)_R\) and include it in each species-density derivative \(n_{s,x}\), following Eqs.~(12)--(15) in Section~2.6 of Xie et al.~\cite{xie2026reduced}. In the sum \(P_x=\sum_s(\theta_s n_{s,x}+n_s\theta_{s,x})\), the coefficient of \(\Phi_x\) is \(-e\sum_s Z_s n_s=0\) by quasineutrality. The species densities \(n_s(R,x)\) that enter \(P_x\) still depend on the electrostatic potential \(\Phi\). Equation~\eqref{eq:2-11} converts \(P_x\) to the derivative with respect to the physical flux \(\psi\) for the pressure contribution to the current source. The normalised-flux coordinate \(x\) and flux difference \(\psi_b-\psi_a\) are obtained from the poloidal-flux distribution at the current equilibrium iterate.

For prescribed external coil currents, the total poloidal flux is

\[
\psi(R,Z)=\psi_{\mathrm{coil}}(R,Z)+\mathcal G[J_\phi](R,Z).
\tag{2.12}
\label{eq:2-12}
\]

Here, \(\mathcal G[J_\phi]\) is the poloidal flux produced by the plasma current, with a free-space Green-function relation applied at the computational boundary. The current occupies the plasma region \(\mathcal A_p\) enclosed by the LCFS and containing the magnetic axis. The magnetic axis and plasma boundary are determined from the total flux. For a diverted equilibrium, the boundary flux equals the flux at the X point on the plasma separatrix. For a limited equilibrium, contact with the prescribed limiter sets the boundary. The flux calculation extends over the surrounding region, including the external separatrix branches.

In the finite-edge-density cases of Section~\ref{sec:4}, \(\Omega(1)=0\). The quasineutral reference densities then give \(P_{\mathrm{MF}}(R,1)=P_{\mathrm{MHD}}(R,1)=P_0(1)\). The prescribed uniform pressure in the current-free exterior equals \(P_0(1)\). The toroidal-field function takes the exterior value \(F(1)=F_{\mathrm{vac}}\) at the interface.

For fixed-total-current calculations, \(\gamma\) is determined together with the current distribution and poloidal-flux distribution so that

\[
\int_{\mathcal A_p}J_\phi\,\mathrm dR\,\mathrm dZ=I_p^{\mathrm{target}}.
\tag{2.13}
\label{eq:2-13}
\]

With \(\gamma\) prescribed, the total current is a solution output.

With the total current fixed, the spatial distributions of \(J_P\) and \(J_F\) can adjust within the coupled solution.

Each multi-fluid and rotating-MHD pair uses the same device configuration, full external coil-current set, reference radius, choice of prescribed total current or prescribed \(\gamma\), and toroidal-field parameterisation. The models share \(P_0(x)\), \(\rho_0(x)\) and \(\Omega(x)\) derived from the species inputs. Equations~\eqref{eq:2-6} and \eqref{eq:2-7} give the rotating-MHD pressure from these profiles, so cases with the same profiles and prescribed equilibrium inputs can share one rotating-MHD reference.

The two models use the same spatial discretisation and checks on the magnetic flux and toroidal current; the MF calculation also checks local quasineutrality. Each model independently determines its pressure, toroidal current, poloidal flux, magnetic axis and plasma boundary, as well as \(\gamma\) in fixed-current mode.

\FloatBarrier
\subsection{Numerical solution method}
\label{sec:2-3}

On a uniform \(R\)--\(Z\) grid, the iterate \(\mathbf i\) contains \(N\) nodal currents \(i_k=J_{\phi,k}\Delta R\Delta Z\) inside the limiter. At each residual evaluation, Eq.~\eqref{eq:2-12} gives the poloidal flux \(\psi(\mathbf i)\). The magnetic axis, plasma boundary and normalised flux are then recalculated, and the total-pressure expression and Eq.~\eqref{eq:2-11} give the model current vector \(\widehat{\mathbf i}\). The discrete fixed-current equations are

\[
\mathbf R_J(\mathbf i,\gamma)
=\mathbf i-\widehat{\mathbf i}\!\left[\psi(\mathbf i),\gamma\right]
=\mathbf 0,\qquad
R_I(\mathbf i)=\sum_{k=1}^{N}i_k-I_p^{\mathrm{target}}=0.
\tag{2.14}
\label{eq:2-14}
\]

In fixed-current mode, \(\mathbf i\) and \(\gamma\) are updated together. When \(\gamma\) is prescribed, the total-current equation is omitted.

The poloidal-flux equation is discretised on a rectangular grid using a fourth-order finite-difference scheme, with a Green-function relation imposed at the computational boundary. It is solved using a discrete sine transform in \(Z\) and banded linear systems in \(R\). Sine-transform approaches to Grad-Shafranov solvers have been described previously \cite{barp2012realtime}. The device and coil interfaces and selected magnetic-field and equilibrium-boundary routines use FreeGS4E/FreeGSNKE components \cite{amorisco2024freegsnke}. The source is evaluated at grid nodes, and critical-point positions are refined by quadratic interpolation. At each multi-fluid source evaluation, we solve the scalar quasineutrality equation for \(\Phi\) and assemble \(P_x\) from the species-density derivatives as described after Eq.~\eqref{eq:2-11}.

The nonlinear equilibrium equations are solved with a matrix-free Newton--Krylov method, using GMRES for the linearised correction~\cite{knoll2004jfnk}. The fixed-current updates in the current-based formulation use a current-preserving null-space representation. Step length is adjusted according to residual reduction. If the model cannot be evaluated at a trial state, the step is reduced and retried.

\begin{figure}[H]
\centering
\includegraphics[width=\linewidth,pagebox=cropbox]{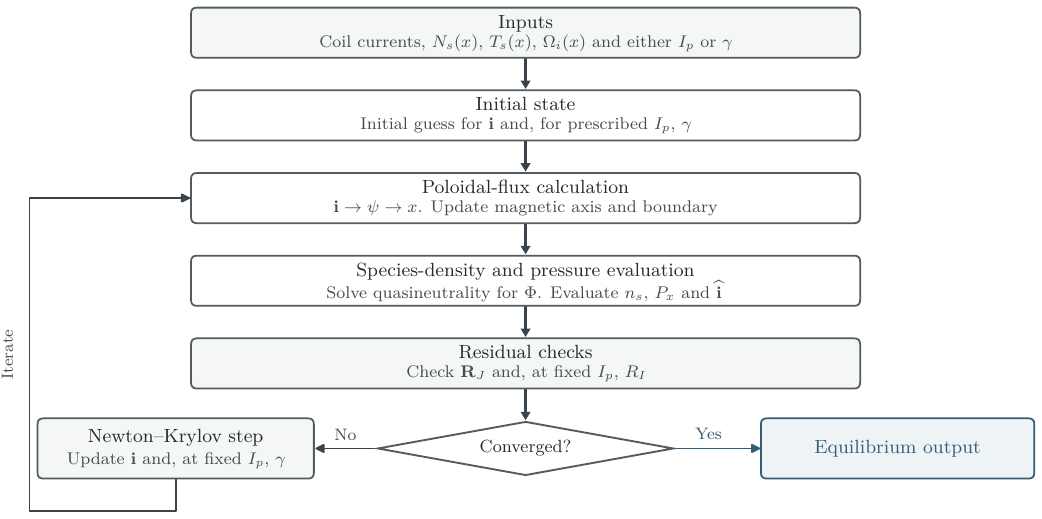}
\caption{Current-based multi-fluid free-boundary iteration. See Appendix~\ref{app:numerical-checks} for the acceptance checks.}
\label{fig:2-1}
\end{figure}

Figure~\ref{fig:2-1} shows the current-based MF iteration.

Acceptance requires a relative source-consistency residual \(r_J\leq10^{-8}\). In fixed-current mode, the total currents obtained from \(\mathbf i\) and \(\widehat{\mathbf i}\) must each be within \(6\ \mathrm A\) of the target. Appendix~\ref{app:numerical-checks} gives the complete criteria.

\FloatBarrier
\section{Numerical verification}
\label{sec:3}

\FloatBarrier
\subsection{MHD limiting relations}
\label{sec:3-1}

At zero rotation, reference quasineutrality gives \(\Phi=0\) and \(n_s=N_s\). The MF pressure and its flux derivative therefore reduce to

\[
P_{\mathrm{MF}}=P_0,\qquad
(\partial_xP_{\mathrm{MF}})_R=P_{0,x}.
\tag{3.1}
\label{eq:3-1}
\]

These are also the static-MHD values and the zero-rotation limit of Eq.~\eqref{eq:2-7}.

For a rotating plasma containing singly charged deuterium and electrons, \(N_D=N_e\) and quasineutrality requires \(n_D=n_e\). Applying these conditions to Eq.~\eqref{eq:2-2} gives

\[
e\Phi=\frac{\theta_e}{\theta_D+\theta_e}W_D,\qquad
n_D=N_D\exp\!\left(\frac{W_D}{\theta_D+\theta_e}\right).
\tag{3.2}
\label{eq:3-4}
\]

With \(P_0=N_D(\theta_D+\theta_e)\), \(\rho_0=m_DN_D\) and \(\Omega=\Omega_D\), the pressure becomes

\[
P_{\mathrm{MF}}=n_D(\theta_D+\theta_e)
=P_0\exp\!\left[\frac{\rho_0\Omega^2(R^2-R_0^2)}{2P_0}\right]
=P_{\mathrm{MHD}}.
\tag{3.3}
\label{eq:3-6}
\]

The pressure expressions coincide for all \(R,x\), so their derivatives \(P_x\) also agree. This recovers the single-ion limit of the reduced model~\cite{xie2026reduced}.

\FloatBarrier
\subsection{MHD-limit tests and pure-DT benchmark}
\label{sec:3-2}

Numerical comparisons are performed in MAST-U-like and ITER-like diverted configurations. For each device, the zero-rotation DT--He equilibrium is calculated with the multi-fluid, rotating-MHD and static-MHD models. The finite-rotation single-ion equilibrium contains deuterium ions and electrons and is calculated with the multi-fluid and rotating-MHD models. The static-MHD reference uses \(P=P_0(x)\), and the rotating-MHD reference evaluates Eq. \eqref{eq:2-7} directly from the macroscopic profiles in Eq. \eqref{eq:2-6}. All three models use the spatial discretisation and numerical solution method described in Section 2.3. Within each pair, the models use the same prescribed coil currents, reference profiles, total plasma current and toroidal-field parameterisation. Each model determines \(\gamma\) in its own fixed-current equilibrium solve.

The MAST-U-like cases use \(R_0=0.900\ \mathrm m\), \(I_p=0.600\ \mathrm{MA}\), \(N_{e0}=1.20\times10^{19}\ \mathrm{m^{-3}}\) and \(T_{e0}=T_{i0}=1.50\ \mathrm{keV}\). The ITER-like cases use \(R_0=6.20\ \mathrm m\), \(I_p=10\ \mathrm{MA}\), \(N_{e0}=5.00\times10^{19}\ \mathrm{m^{-3}}\) and \(T_{e0}=T_{i0}=8.00\ \mathrm{keV}\). The DT--He reference densities satisfy \(N_D=N_T=0.475N_e\) and \(N_{\mathrm{He}}=0.0250N_e\), with \(Z_{\mathrm{He}}=2\).

The reference-density profiles in Eq.~\eqref{eq:2-4} have zero edge values and \(\beta_N=1\), with \(\alpha_N=1\) for the MAST-U-like cases and \(\alpha_N=2\) for the ITER-like cases. The temperature profiles have \(\alpha_T=\beta_T=1\) and edge values equal to \(10^{-3}\) of their axis values. The finite-rotation profiles have \(\alpha_\Omega=2\), \(\beta_\Omega=4\) and zero edge values. The single-ion cases use \(M_{D,\mathrm{ref}}=1.50\) in both devices. In Eq. \eqref{eq:2-5}, \(\alpha_F=2\) and \(\beta_F=1\), with \(F_{\mathrm{vac}}=0.500\ \mathrm{T\,m}\) for MAST-U-like cases and \(32.9\ \mathrm{T\,m}\) for ITER-like cases.

All limiting comparisons use \(129\times257\) grids. The native computational domain is \(0.100\leq R\leq2.0\ \mathrm m\), \(-2.20\leq Z\leq2.2\ \mathrm m\) for MAST-U-like cases and \(3.20\leq R\leq8.8\ \mathrm m\), \(-5.00\leq Z\leq5.0\ \mathrm m\) for ITER-like cases. For the full-device comparison, flux outside each native domain is reconstructed from the prescribed coil currents and the equilibrium plasma current of each model using Green-function sums. The field differences are measured by

\[
\epsilon_\psi=
\frac{\|\psi_{\mathrm{pl}}-\psi_{\mathrm{pl,ref}}\|_2}
{\|\psi_{\mathrm{pl,ref}}\|_2},\qquad
\epsilon_J=
\frac{\|J_\phi-J_{\phi,\mathrm{ref}}\|_2}
{\|J_{\phi,\mathrm{ref}}\|_2},
\tag{3.4}
\label{eq:3-7}
\]

where \(\psi_{\mathrm{pl}}=\psi-\psi_{\mathrm{coil}}\) is the plasma-generated poloidal flux, \(J_\phi\) is the toroidal current density, and the unweighted discrete \(L_2\) norms are evaluated on the common computational grid. The reference is static MHD for the zero-rotation comparisons and rotating MHD for the single-ion comparisons.

For \(d_A\) and \(d_X\), we locate the magnetic axis and the X point that determines the boundary by quadratic interpolation as described in Section~\ref{sec:2-3}. To trace the LCFS, we use that X-point estimate to refine the saddle position on the bicubic interpolant of the total flux and trace the contour at the refined saddle flux. The geometric distances are

\[
\begin{aligned}
d_A&=\|\mathbf a-\mathbf a_{\mathrm{ref}}\|_2,\qquad
d_X=\|\mathbf X-\mathbf X_{\mathrm{ref}}\|_2,\\
d_B(\mathcal C_1,\mathcal C_2)
&=\max\!\left\{
\sup_{\mathbf p\in\mathcal C_1}\inf_{\mathbf q\in\mathcal C_2}
\|\mathbf p-\mathbf q\|_2,\ 
\sup_{\mathbf q\in\mathcal C_2}\inf_{\mathbf p\in\mathcal C_1}
\|\mathbf q-\mathbf p\|_2
\right\},
\end{aligned}
\tag{3.5}
\label{eq:3-8}
\]

where \(\mathbf a\) and \(\mathbf X\) denote the positions of the magnetic axis and that X point, and \(\mathcal C_1\) and \(\mathcal C_2\) are the two LCFS curves. The symmetric maximum distance \(d_B\) is evaluated by point-to-segment distances on the refined curves. All geometric distances are reported in millimetres.

\begin{figure}[!htbp]
\centering
\includegraphics[width=\linewidth,height=0.78\textheight,keepaspectratio]{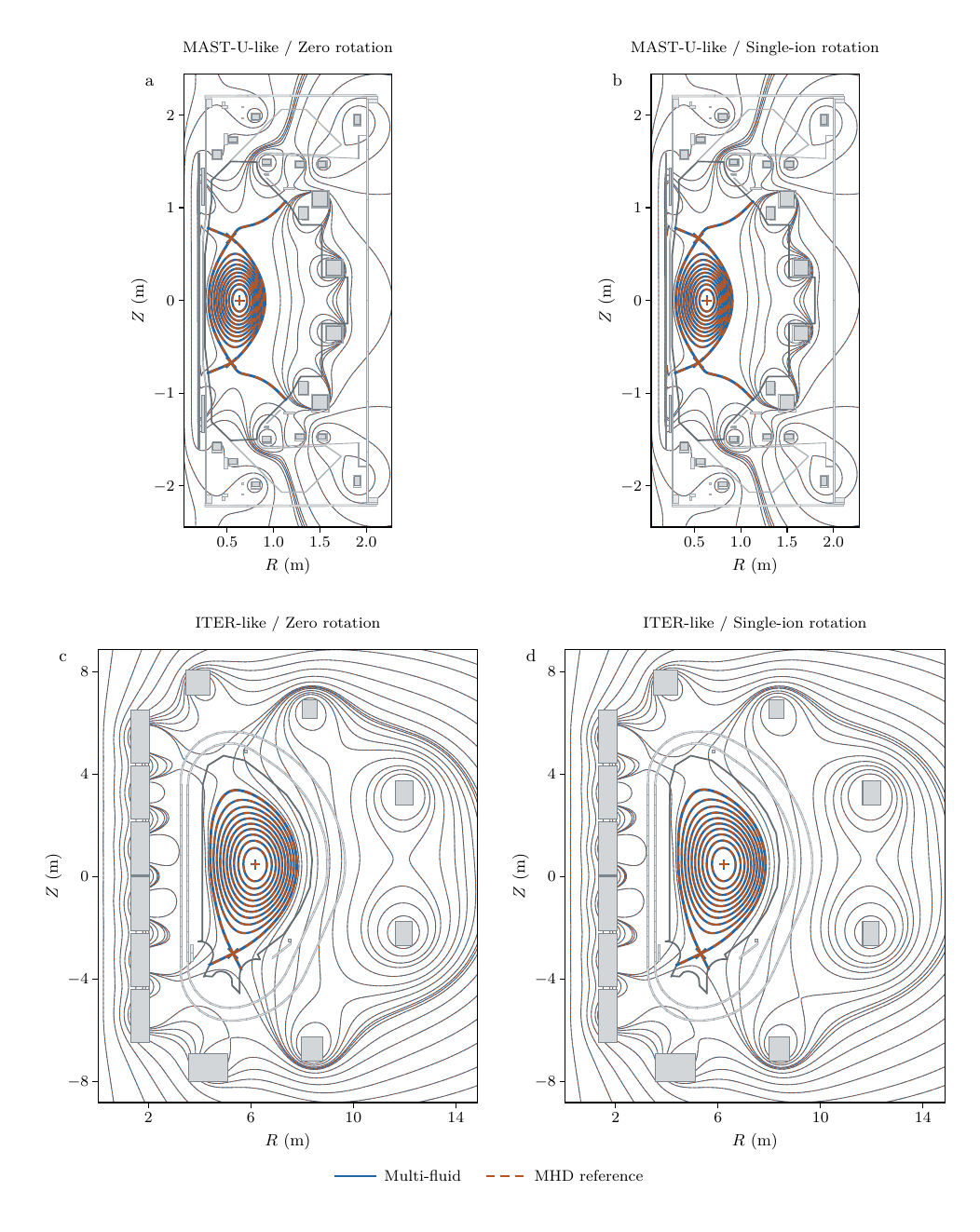}
\caption{\textbf{Magnetic-flux comparisons in the MHD limits.} Panels (a,b) show MAST-U-like configurations and (c,d) ITER-like configurations. The left column compares multi-fluid and static-MHD equilibria for non-rotating DT--He, and the right column compares multi-fluid and rotating-MHD equilibria for rotating deuterium--electron plasmas. Blue solid and orange dashed curves denote multi-fluid and MHD results. Thick curves show internal flux surfaces, the LCFS and divertor branches. Thin curves show contours of the total poloidal flux outside the plasma. Plus and cross symbols mark the magnetic axes and the X points that determine the plasma boundaries, respectively.}
\label{fig:3-1}
\end{figure}

The MF and MHD interior surfaces, LCFS curves and divertor branches coincide in both limits in both devices (Fig.~\ref{fig:3-1}). The exterior flux contours, which include the contributions of the coils and plasma current, also overlap throughout the region shown in the figure. Table~\ref{tab:3-1} also includes the rotating-MHD versus static-MHD comparison at zero rotation. The largest relative differences are \(1.62\times10^{-15}\) for plasma-generated flux and \(6.00\times10^{-15}\) for toroidal current density. The magnetic-axis and X-point coordinates coincide in double precision. LCFS differences are below the reported geometric-extraction thresholds of \(0.0100\ \mathrm{mm}\) for MAST-U-like cases and \(0.00100\ \mathrm{mm}\) for ITER-like cases.

The two model expressions give relative \(L_2\) differences below \(2.10\times10^{-15}\) for both \(P\) and \(P_x\) at common \((R,x)\) points in either limit.

\begin{table}[!htbp]
\centering
\small
\setlength{\tabcolsep}{3pt}
\renewcommand{\arraystretch}{1.12}
\caption{Field and geometry differences in the MHD limits.}
\label{tab:3-1}
\fittable{\begin{tabular}{@{}lllrrrrr@{}}
\toprule
Device & Limit & \shortstack{Model\\reference} & $\epsilon_\psi$ & $\epsilon_J$ & \shortstack{$d_A$\\mm} & \shortstack{$d_X$\\mm} & \shortstack{$d_B$\\mm} \\
\midrule
MAST-U-like & Zero rotation & MF / sMHD & \(4.37\times10^{-16}\) & \(1.19\times10^{-15}\) & 0 & 0 & \(< 0.0100\) \\
MAST-U-like & Zero rotation & rMHD / sMHD & 0 & 0 & 0 & 0 & \(< 0.0100\) \\
ITER-like & Zero rotation & MF / sMHD & \(1.81\times10^{-16}\) & \(2.15\times10^{-15}\) & 0 & 0 & \(< 0.00100\) \\
ITER-like & Zero rotation & rMHD / sMHD & 0 & 0 & 0 & 0 & \(< 0.00100\) \\
MAST-U-like & Single ion & MF / rMHD & \(1.31\times10^{-16}\) & \(2.39\times10^{-15}\) & 0 & 0 & \(< 0.0100\) \\
ITER-like & Single ion & MF / rMHD & \(1.62\times10^{-15}\) & \(6.00\times10^{-15}\) & 0 & 0 & \(< 0.00100\) \\
\bottomrule
\end{tabular}}
\par\smallskip{\footnotesize\raggedright MF, rMHD and sMHD denote multi-fluid, rotating MHD and static MHD. Field and geometry measures are defined in Eqs. \eqref{eq:3-7} and \eqref{eq:3-8}. Zero coordinate distances denote identical double-precision coordinates. The LCFS reporting threshold is obtained by rounding the sum of the final geometric-refinement changes of the two curves upward to a power of ten. Distances below this threshold are shown with a less-than sign.\par}
\end{table}

\FloatBarrier

The rotating pure-DT benchmark uses the \(15\ \mathrm{MA}\) case in Section~\ref{sec:4-3}, with \(\Omega_0=252\ \mathrm{krad\,s^{-1}}\) and \(k=0.50\). For this case, MFEQ-FB iterated on the plasma-generated poloidal flux. At each trial flux, it solved for \(\gamma\) to satisfy the prescribed total-current constraint and corrected the flux with Newton--GMRES. The converged solution satisfies Eq.~\eqref{eq:2-14} and the acceptance criteria in Appendix~\ref{app:numerical-checks}.

This benchmark tests the total-pressure expression and the interior equilibrium beyond the MHD limits. At a given poloidal flux, we compare the numerical potential and pressure with their analytical values. At a given plasma boundary, we compare the interior equilibrium with a separate fixed-boundary calculation.

The quasineutrality condition has a closed form when singly charged deuterium and tritium share \(\theta_i(x)\) and \(\Omega(x)\) and \(W_e=0\) as in Eq.~\eqref{eq:2-2}. At given \(R,x\) and reference profiles, \(n_D+n_T=n_e\) gives
\[
\begin{gathered}
e\Phi=\frac{\theta_e\theta_i}{\theta_e+\theta_i}
\ln\!\left(\frac{N_D\exp(W_D/\theta_i)+N_T\exp(W_T/\theta_i)}{N_e}\right),\\
n_e=N_e\exp(e\Phi/\theta_e),\qquad P_{\mathrm{MF}}=(\theta_e+\theta_i)n_e.
\end{gathered}
\]
For \(Q=\Phi\), \(P_{\mathrm{MF}}\) or \(P_x\), the relative maximum difference over the grid nodes within the LCFS is
\[
\epsilon_Q=\frac{\max|Q_{\mathrm{analytic}}-Q_{\mathrm{stored}}|}{\max|Q_{\mathrm{stored}}|}.
\]
The relative maximum differences between numerical and analytical values are \(1.06\times10^{-14}\) and \(1.07\times10^{-14}\), for the potential and pressure, respectively, over \(44{,}633\) nodes evaluated using the saved total flux. Independently differentiating the analytical pure-DT pressure at fixed \(R\) gives \(\epsilon_{P_x}<4\times10^{-14}\) on the saved grid nodes within the LCFS of each of the three pure-DT temperature-ratio cases in Section~\ref{sec:4-3}.

For the prescribed LCFS, the fixed-boundary MFEQ calculation gives nearly the same interior flux and toroidal current density as MFEQ-FB. Over the \(44{,}633\) common plasma nodes, the relative \(L_2\) differences in \(u=\psi-\psi_b\) and \(J_\phi\) are \(0.0577\%\) and \(0.0583\%\), respectively. These norms use poloidal cell-area weights within the prescribed LCFS and are normalised by the corresponding MFEQ-FB field norms. On the common \(257\times513\) physical grid, the magnetic axes are \(0.21\ \mathrm{mm}\) apart. On the \(x=0.50\) surface at \(Z=0.50\ \mathrm m\), the T/D separation factors are \(3.33707\) for MFEQ and \(3.33704\) for MFEQ-FB. For this lowest-temperature-ratio pure-DT case (\(k=0.50\)), the adapted MFEQ code used the LCFS from MFEQ-FB and the same physical inputs. Starting from an independent initial guess, it solved for the interior poloidal flux and determined \(\gamma\) for the prescribed current of \(15\ \mathrm{MA}\).

\subsection{Grid convergence analysis}
\label{sec:3-3}

Grid refinement is examined for rotating ITER-like DT and DT--He equilibria. The DT reference densities satisfy \(N_D=N_T=0.500N_e\), and the DT--He composition is the same as in Section 3.2. Both plasmas use the ITER-like density, temperature, toroidal-field inputs and \(10\ \mathrm{MA}\) current constraint specified there, with the common ion rotation increased to \(M_{D,\mathrm{ref}}=2.35\). The finest-grid multi-fluid equilibria have \(M_{D,\mathrm{axis}}\simeq2.46\) for both plasmas. Each plasma is calculated with the multi-fluid and rotating-MHD models on \(129\times257\), \(193\times385\) and \(257\times513\) grids over the same computational domain. The corresponding maximum mesh spacings are \(43.8\), \(29.2\) and \(21.9\ \mathrm{mm}\). Physical profiles, prescribed coil currents, current constraint and toroidal-field parameterisation are fixed across grids, with \(\gamma\) determined in each equilibrium solve.

\begin{figure}[H]
\centering
\captionsetup{skip=5pt}
\includegraphics[width=\linewidth,height=.455\textheight,keepaspectratio,pagebox=cropbox]{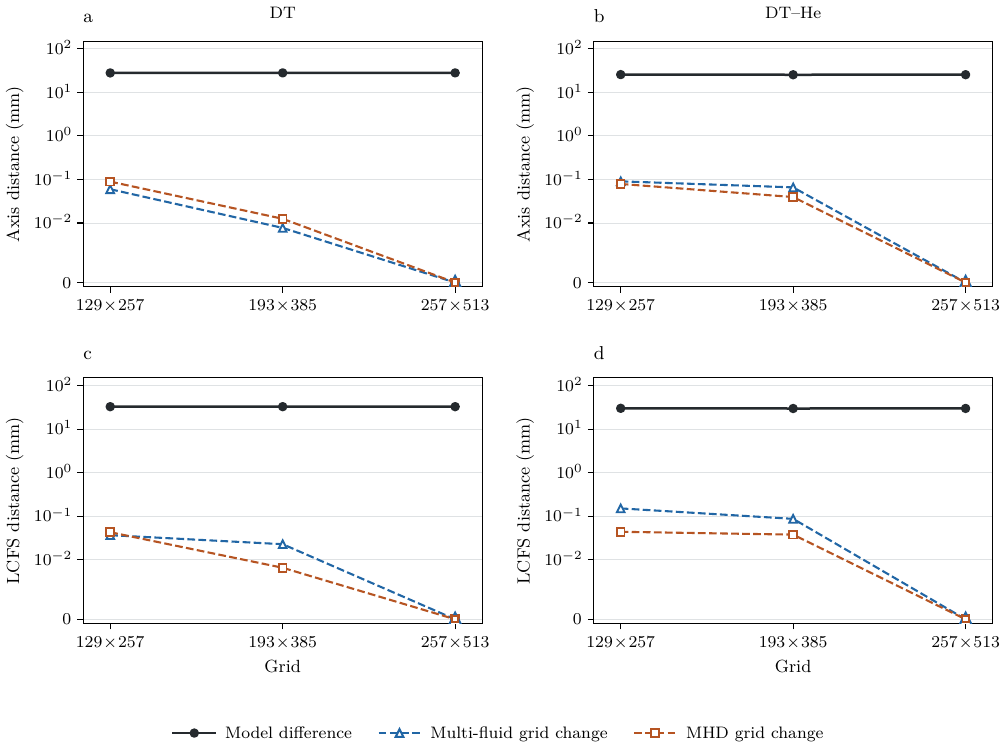}
\caption{Grid refinement in rotating ITER-like DT (a,c) and DT--He (b,d). Circles show the MF--MHD axis distance in (a,b) and the symmetric maximum LCFS distance in (c,d). Triangles and squares give MF and rotating-MHD changes from their own \(257\times513\) solutions. The ordinate is linear below \(0.00200\ \mathrm{mm}\) and logarithmic above.}
\label{fig:3-2}
\end{figure}

\begin{samepage}
Let \(h_f\) denote the finest grid. For each model \(m\), the solution on \(h_f\) provides the reference for changes in the plasma-generated flux, magnetic-axis position and LCFS geometry under grid refinement. The model difference is evaluated separately on each grid.\par
\nopagebreak
With \(\mathcal I_h\) denoting bicubic interpolation from the finest grid to the current grid, the measures are
\[
\begin{aligned}
\epsilon_{\psi,h}^{(m)}
&=\frac{\|\psi_{\mathrm{pl},h}^{(m)}-\mathcal I_h\psi_{\mathrm{pl},h_f}^{(m)}\|_2}
{\|\mathcal I_h\psi_{\mathrm{pl},h_f}^{(m)}\|_2},\\
d_{A,h}^{(m)}&=\|\mathbf a_h^{(m)}-\mathbf a_{h_f}^{(m)}\|_2,\\
d_{B,h}^{(m)}&=d_B(\mathcal C_h^{(m)},\mathcal C_{h_f}^{(m)}),\\
\Delta_A(h)&=\|\mathbf a_h^{(\mathrm{MF})}-\mathbf a_h^{(\mathrm{MHD})}\|_2,\\
\Delta_B(h)&=d_B(\mathcal C_h^{(\mathrm{MF})},\mathcal C_h^{(\mathrm{MHD})}),
\end{aligned}
\tag{3.6}
\label{eq:3-9}
\]

where MHD denotes the rotating-MHD reference in this section. The same geometric extraction and distance definitions are used for both models and all grids. Figure~\ref{fig:3-2} shows the resulting axis and LCFS distances.
\par
\end{samepage}

\FloatBarrier

At \(10\ \mathrm{MA}\), the MF--rotating-MHD differences in magnetic-axis and LCFS positions exceed the measured geometric changes under refinement by more than two orders of magnitude (Fig.~\ref{fig:3-2} and Table~\ref{tab:3-2}). On the finest grid, the axis and LCFS differences are \(28.0\ \mathrm{mm}\) and \(32.7\ \mathrm{mm}\) for DT, and \(25.4\ \mathrm{mm}\) and \(29.8\ \mathrm{mm}\) for DT--He. Across the two coarser grids, the largest changes relative to the finest grid are \(0.0907\ \mathrm{mm}\) for the axis and \(0.151\ \mathrm{mm}\) for the LCFS. These distances decrease for both models in both mixtures from \(129\times257\) to \(193\times385\). For rotating MHD in DT--He, the same refinement increases the relative plasma-flux difference \(\epsilon_{\psi,h}\) from \(5.48\times10^{-6}\) to \(8.21\times10^{-6}\).

\begin{table}[H]
\centering
\footnotesize
\setlength{\tabcolsep}{3pt}
\renewcommand{\arraystretch}{1.12}
\caption{Flux and geometry changes with grid refinement for rotating ITER-like DT and DT--He equilibria.}
\label{tab:3-2}
\fittable{\begin{tabular}{@{}llrrrrrrrr@{}}
\toprule
Plasma & Grid & $\epsilon_{\psi,h}^{MF}$ & $\epsilon_{\psi,h}^{MHD}$ & $d_{A,h}^{MF}$ & $d_{A,h}^{MHD}$ & $d_{B,h}^{MF}$ & $d_{B,h}^{MHD}$ & $\Delta_A$ & $\Delta_B$ \\
\midrule
DT & \(129\times257\) & \(4.10\times10^{-6}\) & \(5.96\times10^{-6}\) & 0.0598 & 0.0893 & 0.0367 & 0.0429 & 27.9 & 32.7 \\
DT & \(193\times385\) & \(2.84\times10^{-6}\) & \(9.04\times10^{-7}\) & 0.00774 & 0.0125 & 0.0228 & 0.00656 & 28.0 & 32.7 \\
DT & \(257\times513\) & 0 & 0 & 0 & 0 & 0 & 0 & 28.0 & 32.7 \\
DT--He & \(129\times257\) & \(3.41\times10^{-5}\) & \(5.48\times10^{-6}\) & 0.0907 & 0.0790 & 0.151 & 0.0439 & 25.5 & 30.0 \\
DT--He & \(193\times385\) & \(1.90\times10^{-5}\) & \(8.21\times10^{-6}\) & 0.0659 & 0.0395 & 0.0870 & 0.0378 & 25.3 & 29.7 \\
DT--He & \(257\times513\) & 0 & 0 & 0 & 0 & 0 & 0 & 25.4 & 29.8 \\
\bottomrule
\end{tabular}}
\par\smallskip{\raggedright All geometric distances are in millimetres. The measures are defined in Eq. \eqref{eq:3-9}, with each model's \(257\times513\) solution as its grid reference.\par}
\end{table}

The MHD-limit tests and grid-refinement calculations meet the acceptance criteria in Appendix~\ref{app:numerical-checks}. Table~\ref{tab:acceptance-residuals} reports the residuals and current errors of the grid-refinement calculations.

\begin{samepage}
The preceding grid-refinement study uses \(10\ \mathrm{MA}\) inputs. Table~\ref{tab:3-3} reports two-grid changes in selected results from the \(15\ \mathrm{MA}\) application scans with finite edge density in Sections~\ref{sec:4-2} and \ref{sec:4-3}. The rotating MHD comparisons use the common equilibrium computed on each grid. The multiplier \(k\) of the temperature-ratio profile is defined in Eq.~\eqref{eq:4-4}.

\begin{table}[H]
\centering
\small
\setlength{\tabcolsep}{7pt}
\renewcommand{\arraystretch}{1.13}
\caption{Two-grid comparison for selected \(15\ \mathrm{MA}\) application cases at \(\Omega_0=252\ \mathrm{krad\,s^{-1}}\).}
\label{tab:3-3}
\begin{tabular}{@{}>{\raggedright\arraybackslash}p{.57\linewidth}rrr@{}}
\toprule
Observable and reference & \(193\times385\) & \(257\times513\) & \(\lvert\Delta_h\rvert\) \\
\midrule
\multicolumn{4}{@{}l}{\textit{MF outboard crossing shifts on \(x=0.50\) (mm)}} \\
\quad \(2.5\%\) He relative to pure DT, \(k=1.00\) & \(-5.17\) & \(-8.38\) & \(3.21\) \\
\quad \(5\%\) He relative to pure DT, \(k=1.00\) & \(-11.8\) & \(-15.5\) & \(3.73\) \\
\quad \(5\%\) He, \(k=1.20\) relative to \(k=1.00\) & \(-9.35\) & \(-6.12\) & \(3.23\) \\
\quad Pure DT, \(k=1.20\) relative to \(k=1.00\) & \(-7.17\) & \(-10.7\) & \(3.56\) \\
\addlinespace[3pt]
\multicolumn{4}{@{}l}{\textit{Pure DT, MF relative to rotating MHD (mm)}} \\
\quad LCFS crossing shift, \(k=0.50\) (MF minus MHD) & \(+143.2\) & \(+142.6\) & \(0.566\) \\
\quad Full-contour LCFS RMS distance, \(k=0.50\) & \(84.4\) & \(84.0\) & \(0.370\) \\
\quad Full-contour LCFS RMS distance, \(k=1.20\) & \(36.9\) & \(36.3\) & \(0.698\) \\
\addlinespace[3pt]
MF \(S_{T/D}\), pure DT, \(k=0.50\) (dimensionless) & \(3.339\) & \(3.337\) & \(0.00158\) \\
\bottomrule
\end{tabular}
\par\smallskip{\footnotesize\raggedright Crossing shifts use the outboard intersections at \(Z=0.50\ \mathrm m\). Negative and positive values indicate inward and outward shifts, respectively. The separation factor uses both crossings of the \(x=0.50\) surface on that cut. The last column gives the observed absolute change upon refinement, computed before rounding.\par}
\end{table}
\end{samepage}

All four selected \(x=0.50\) surface shifts remain inward on both grids, with magnitudes changing by \(3\)--\(4\ \mathrm{mm}\) upon refinement (Table~\ref{tab:3-3}). In pure DT, the full-LCFS RMS distances at \(k=0.50\) and \(1.20\), and the LCFS crossing shift at \(k=0.50\), each change by less than \(1\ \mathrm{mm}\).

\FloatBarrier
\section{Numerical applications}
\label{sec:4}

Using the ITER-like machine geometry distributed with FreeGSNKE~\cite{amorisco2024freegsnke}, we compare rotation-induced changes in pure DT and changes in species inputs under a common rotating MHD reference. Section~\ref{sec:4-1} includes a separate \(11\ \mathrm{MA}\) configuration. The composition and temperature-ratio scans hold the coil currents, total plasma current (\(15\ \mathrm{MA}\)) and reference profiles \(N_e(x)\), \(\rho_0(x)\), \(P_0(x)\) and \(\Omega(x)\) fixed, with \(\Omega_0=252\ \mathrm{krad\,s^{-1}}\). All cases in these two scans therefore share one computed rotating MHD reference equilibrium per grid. The scans test how ion composition and the partition of thermal pressure between ions and electrons affect the self-consistent MF boundary under these constraints.

The \(15\ \mathrm{MA}\) scans use a common set of coil currents, \(R_0=6.20\ \mathrm m\), \(F_{\mathrm{vac}}=32.86\ \mathrm{T\,m}\) and \((\alpha_F,\beta_F)=(2,1)\) in Eq.~\eqref{eq:2-5}. The uniform pressure prescribed in the current-free exterior equals the LCFS thermal pressure of \(20.96\ \mathrm{kPa}\). The \(415\ \mathrm s\) snapshot of the ITPA TC-33 ITER reference case~\cite{bourdelle2025tc33} informs the density and temperature scales. The reported results use \(257\times513\) grids on \(3.2\le R\le8.8\ \mathrm m\), \(-5.0\le Z\le5.0\ \mathrm m\).

\begin{figure}[!htbp]
\centering
\includegraphics[width=\linewidth]{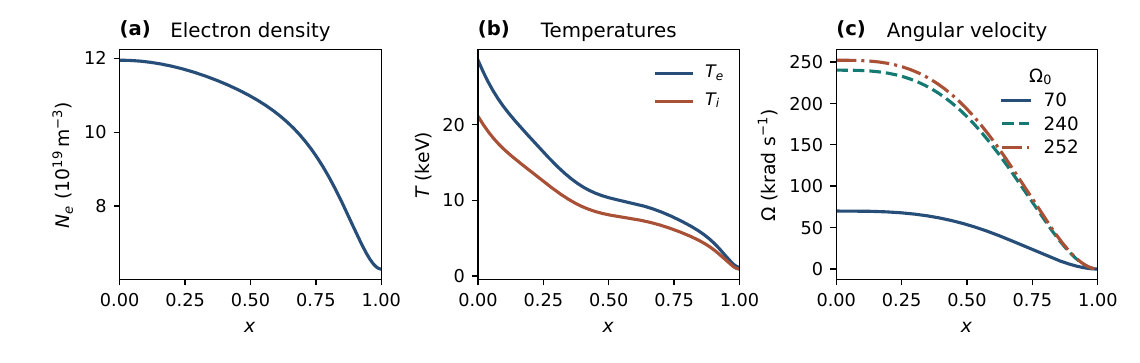}
\caption{Pure-DT reference electron density (a), electron and common ion temperatures (b), and ion angular velocity for the three nonzero amplitudes (c). The nonrotating reference has \(\Omega(x)=0\) throughout.}
\label{fig:4-1}
\end{figure}

Figure~\ref{fig:4-1} shows the pure-DT reference electron density, electron and common ion temperatures, and the three nonzero angular-velocity profiles as functions of the normalised poloidal flux \(x\). The ions are singly charged with \(N_D=N_T=N_e/2\). The prescribed \(N_e(x)\) decreases from \(1.19\times10^{20}\ \mathrm{m^{-3}}\) at \(x=0\) to \(6.30\times10^{19}\ \mathrm{m^{-3}}\) at \(x=1\). The on-axis \(T_e\) and \(T_i\) are \(28.5\) and \(21.1\ \mathrm{keV}\). We vary the amplitude in the profile
\[
\Omega(x)=\Omega_0(1-x^3)^2 .
\tag{4.1}
\label{eq:4-1}
\]
We use \(\Omega_0=0,\ 70,\ 240,\ 252\ \mathrm{krad\,s^{-1}}\), with zero rotation as the static reference. In Budny's \(15\ \mathrm{MA}\) ITER H-mode calculation with three neutral beams, ion-cyclotron heating and GLF23 momentum transport, the near-axis angular velocity is about \(73\ \mathrm{krad\,s^{-1}}\)~\cite{budny2009iter}. Tala et al. obtain about \(70\ \mathrm{krad\,s^{-1}}\) at the centre of the ITER baseline scenario with an imposed pedestal-top value of \(15\ \mathrm{krad\,s^{-1}}\)~\cite{tala2011pinch}. These predictions motivate the \(70\ \mathrm{krad\,s^{-1}}\) amplitude in our prescribed profiles.

Chapman et al.~\cite{chapman2011shear} considered strongly rotating equilibria with ITER poloidal-field coils and on-axis speeds around one quarter of the Alfv\'en speed. At \(240\) and \(252\ \mathrm{krad\,s^{-1}}\), our \(R_0|\Omega_0|/v_{A,\mathrm{ref}}\) is about \(0.22\) and \(0.23\), respectively, using \(5.30\ \mathrm T\) and the on-axis reference mass density of pure DT. We use these two amplitudes as controlled strong-rotation cases.

Table~\ref{tab:4-cases} distinguishes the reference speed \(R_0|\Omega_0|\) from the toroidal speed at the MF magnetic axis, \(v_{\mathrm{ax}}=R_{\mathrm{ax}}|\Omega_0|\). The reference thermal Mach numbers use \(R_0|\Omega_0|/\sqrt{\theta_s(0)/m_s}\) [Eq.~\eqref{eq:2-8}]. The two models use the same angular-velocity profile at each rotation amplitude. At \(70\ \mathrm{krad\,s^{-1}}\), the reference speed is \(434\ \mathrm{km\,s^{-1}}\), and both models give a toroidal speed of approximately \(415\ \mathrm{km\,s^{-1}}\) at their respective magnetic axes.

\begin{table}[!htbp]
\centering
\caption{Rotation and composition cases at \(15\ \mathrm{MA}\) and the baseline temperature ratio. \(v_{\mathrm{ax}}\) is the ion toroidal speed at the MF magnetic axis.}
\label{tab:4-cases}
\small
\setlength{\tabcolsep}{5pt}
\begin{tabular}{@{}lrrrrrr@{}}
\toprule
Mixture & \shortstack{\(\Omega_0\)\\ (\(\mathrm{krad\,s^{-1}}\))} & \shortstack{\(R_0|\Omega_0|\)\\ (\(\mathrm{km\,s^{-1}}\))} & \shortstack{\(v_{\mathrm{ax}}\)\\ (\(\mathrm{km\,s^{-1}}\))} & \(M_{D,\mathrm{ref}}\) & \(M_{T,\mathrm{ref}}\) & \(M_{\mathrm{He},\mathrm{ref}}\) \\
\midrule
DT & 0 & 0 & 0 & 0 & 0 & \textemdash \\
DT & 70 & 434 & 415 & 0.43 & 0.53 & \textemdash \\
DT & 240 & 1490 & 1530 & 1.48 & 1.81 & \textemdash \\
DT & 252 & 1560 & 1660 & 1.55 & 1.90 & \textemdash \\
DT--He (2.5\%) & 252 & 1560 & 1660 & 1.55 & 1.89 & 2.18 \\
DT--He (5\%) & 252 & 1560 & 1660 & 1.54 & 1.88 & 2.17 \\
\bottomrule
\end{tabular}
\end{table}

\FloatBarrier
\subsection{Rotation effects in DT plasmas}
\label{sec:4-1}

\begin{figure}[!htbp]
\centering
\includegraphics[width=\linewidth]{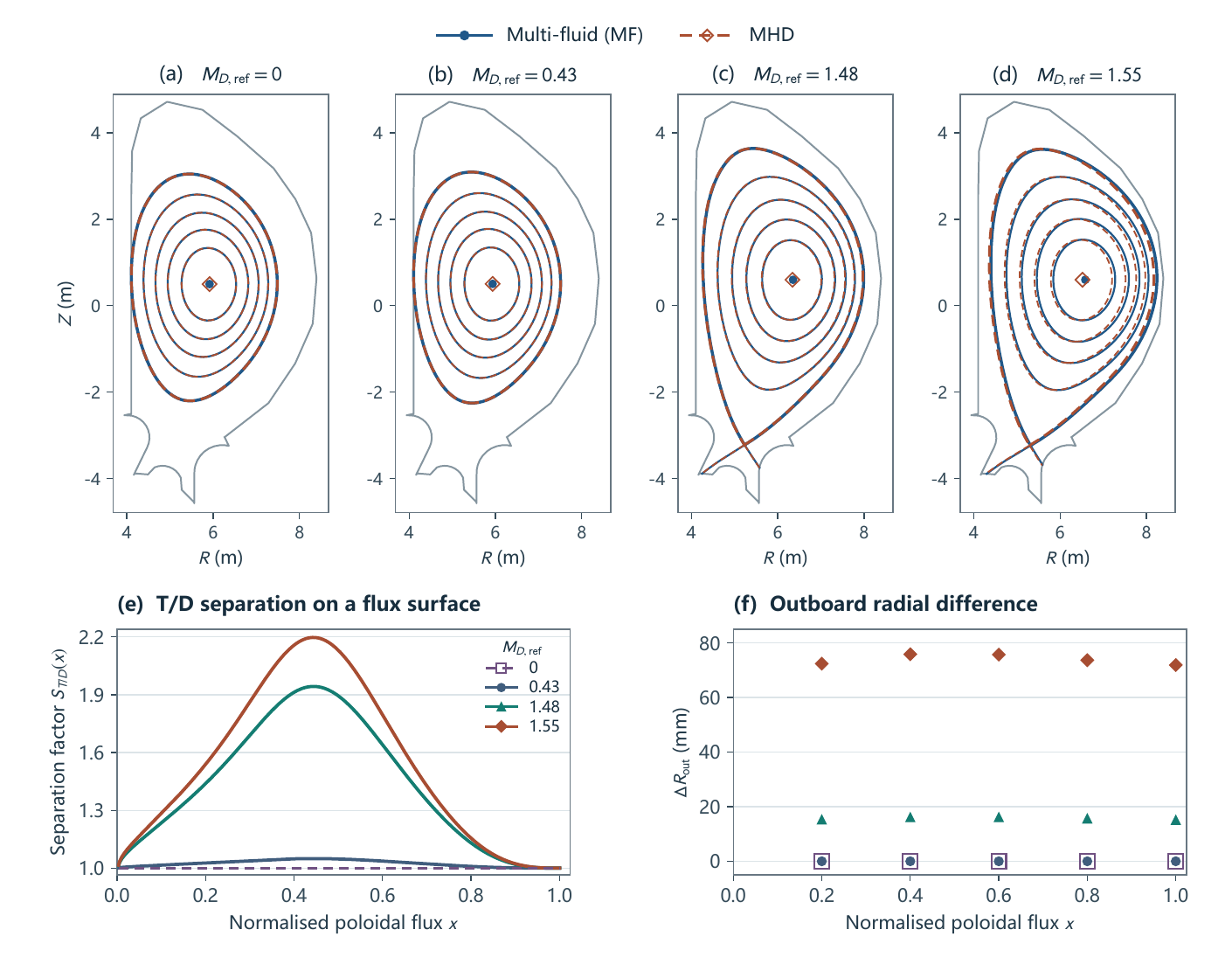}
\caption{Rotation scan in pure DT. Panels (a--d) show MF (solid) and rotating MHD (dashed) flux surfaces. Filled circles and open diamonds mark their respective magnetic axes. On the \(Z_*=0.50\ \mathrm m\) cut, (e) shows the MF separation factor and (f) the outboard crossing radius in MF minus that in rotating MHD at the same flux label.}
\label{fig:4-2}
\end{figure}

The \(15\ \mathrm{MA}\) pure-DT rotation scan uses the baseline reference density and temperature profiles. The on-axis angular velocities \(0,\ 70,\ 240,\ 252\ \mathrm{krad\,s^{-1}}\) correspond to \(M_{D,\mathrm{ref}}=0,\ 0.43,\ 1.48,\ 1.55\). Each panel of Fig.~\ref{fig:4-2}(a--d) pairs MF and rotating MHD at the same prescribed \(\Omega(x)\).

The \(0\) and \(70\ \mathrm{krad\,s^{-1}}\) equilibria are limited and have similar overall contours, while the two faster cases are lower single-null [Fig.~\ref{fig:4-2}(a--d)]. On the \(Z=0.50\ \mathrm m\) cut, the outboard MF LCFS crossing at \(70\ \mathrm{krad\,s^{-1}}\) lies \(24\ \mathrm{mm}\) farther out than in the zero-rotation MF equilibrium with the same remaining inputs.

On the \(Z=0.50\ \mathrm m\) cut, the tritium-to-deuterium density ratios at the outboard and inboard crossings of an MF magnetic surface with label \(x\) give the D--T separation factor
\[
S_{T/D}(x)=
\frac{(n_T/n_D)(R_{\mathrm{out}}(x),x)}
     {(n_T/n_D)(R_{\mathrm{in}}(x),x)} .
\tag{4.2}
\label{eq:4-2}
\]
\begin{samepage}
D and T have the same charge and share \(\theta_i(x)\) and \(\Omega(x)\), so the electrostatic factor in Eq.~\eqref{eq:2-2} cancels in their local density ratio. The centrifugal exponent retains the mass difference,
\[
\frac{n_T(R,x)}{n_D(R,x)}
=\frac{N_T(x)}{N_D(x)}
\exp\!\left[\frac{(m_T-m_D)\Omega^2(x)(R^2-R_0^2)}
                 {2\theta_i(x)}\right] .
\tag{4.3}
\label{eq:4-3}
\]
The local \(n_T/n_D\) therefore rises with \(R\) on a rotating surface. Taking the outboard-to-inboard ratio of Eq.~\eqref{eq:4-3} cancels \(N_T/N_D\) and the \(R_0^2\) term, giving
\[
\ln S_{T/D}(x)=
\frac{(m_T-m_D)\Omega^2(x)}{2\theta_i(x)}
\bigl[R_{\mathrm{out}}^2(x)-R_{\mathrm{in}}^2(x)\bigr] .
\]
The two radii lie on the same magnetic surface and the same cut. With \(\Omega=0\), the exponent vanishes and \(S_{T/D}=1\) wherever the cut crosses the surface twice.
\par
\end{samepage}

For each prescribed angular-velocity profile \(\Omega(x)\), we take the outboard crossing of each model's surface with flux label \(x\) at \(Z_*=0.50\ \mathrm m\) and define
\[
\Delta R_{\mathrm{out}}(x)=R_{\mathrm{out}}^{\mathrm{MF}}(x;Z_*)-
R_{\mathrm{out}}^{\mathrm{MHD}}(x;Z_*) .
\]
Positive values place the MF intersection at larger \(R\). For each Mach number, Fig.~\ref{fig:4-2}(f) marks \(\Delta R_{\mathrm{out}}\) at the four interior surfaces \(x=0.20,\ 0.40,\ 0.60,\ 0.80\) and the LCFS at \(x=1\) shown in panels (a--d).

At zero rotation the two models agree to numerical precision. To compute the symmetric LCFS root-mean-square (RMS) distance, we sample 4096 points uniformly in arc length on each contour and measure each point's distance to the nearest segment of the other contour. We take the square root of the mean squared distance over both directions.

\begin{samepage}
At \(70\ \mathrm{krad\,s^{-1}}\), the MF and rotating MHD boundaries nearly coincide, while the outboard T/D density ratio on the \(x=0.50\) MF surface is \(4.78\%\) higher than the inboard value on the \(Z=0.50\ \mathrm m\) cut [Fig.~\ref{fig:4-2}(e)]. The MF--MHD axis and full LCFS distances are smaller than the corresponding changes within either model from the \(193\times385\) to the \(257\times513\) grid. The same holds for the magnitudes of the outboard-radius differences at \(x=0.50\) and at the LCFS.
\par
\end{samepage}

The two total-pressure expressions have the same second-order centrifugal correction at weak rotation. Scale \(\Omega(x)\) by \(\lambda\) at fixed \(R\), \(x\) and reference profiles. Quasineutrality gives \(\Phi=O(\lambda^2)\) as \(\lambda\) tends to zero. The terms linear in \(\Phi\) cancel in the MF pressure sum because \(\sum_s Z_sN_s=0\) [Eqs.~\eqref{eq:2-2} and \eqref{eq:2-3}]. Both expressions then give \(P_0+\tfrac12\rho_0(x)\lambda^2\Omega^2(x)(R^2-R_0^2)+O(\lambda^4)\) [Eq.~\eqref{eq:2-7} for rotating MHD].

At the same \(R\), \(x\) and reference profiles, \(P_{\mathrm{MF}}\geq P_{\mathrm{MHD}}\). Exponential convexity and reference quasineutrality give this bound for mixtures with helium as well. Holding \(R\), \(x\) and the reference density and temperature profiles fixed, \(P_{\mathrm{MF}}/P_{\mathrm{MHD}}\) is nondecreasing with the common ion angular speed \(|\Omega|\) (Appendix~\ref{app:pressure-ratio}).

\begin{samepage}
For pure DT with \(N_D=N_T=N_e/2\) and a common ion temperature and angular velocity, the exact pressure ratio is
\[
\frac{P_{\mathrm{MF}}}{P_{\mathrm{MHD}}}
=[\cosh\chi(R,x)]^{\theta_i(x)/[\theta_i(x)+\theta_e(x)]},
\qquad
\chi(R,x)=\frac{(m_T-m_D)\Omega^2(x)(R^2-R_0^2)}{4\theta_i(x)}.
\]
\par
\end{samepage}
At fixed geometry, D--T separation first appears at order \(\lambda^2\) and the pressure difference at order \(\lambda^4\) when \(\Omega\) is scaled by \(\lambda\). Rotating MHD includes the centrifugal response to the mean ion mass. The D--T mass difference gives the local T/D density ratio \(\exp(2\chi)\) and the \(\cosh\chi\) factor in the pressure ratio, whose leading correction is proportional to \(\chi^2\). This accounts for the discernible separation and nearly equal pressures at \(70\ \mathrm{krad\,s^{-1}}\).

At \(70\ \mathrm{krad\,s^{-1}}\), the MF pressure is \(0.0044\%\) higher inboard and \(0.0019\%\) higher outboard than the corresponding rotating-MHD pressure. Both expressions are evaluated at \(x=0.50\) and at the radius of each MF crossing on the \(Z=0.50\ \mathrm m\) cut.

Rotating MHD captures most of the outward boundary displacement at \(240\) and \(252\ \mathrm{krad\,s^{-1}}\), while the additional MF displacement grows between these two cases. On the \(Z=0.50\ \mathrm m\) cut, the outboard MF crossing moves outward by \(502\) and \(756\ \mathrm{mm}\), respectively, relative to the zero-rotation MF equilibrium with the other inputs fixed. At the same prescribed \(\Omega(x)\) in each pair, the MF crossing lies about \(15\) and \(72\ \mathrm{mm}\) farther out than the rotating MHD crossing [Fig.~\ref{fig:4-2}(f)].

As \(M_{D,\mathrm{ref}}\) increases from \(1.48\) to \(1.55\), the T/D separation factor on the \(x=0.50\) MF surface increases from \(1.89\) to \(2.13\) on the \(Z=0.50\ \mathrm m\) cut. In both cases separation is strongest on intermediate surfaces and weakens towards the edge [Fig.~\ref{fig:4-2}(e)]. Over the same two cases, the MF--rotating-MHD full-LCFS RMS distance increases from \(8.8\) to \(42.2\ \mathrm{mm}\).

\begin{figure}[!htbp]
\centering
\includegraphics[width=\linewidth]{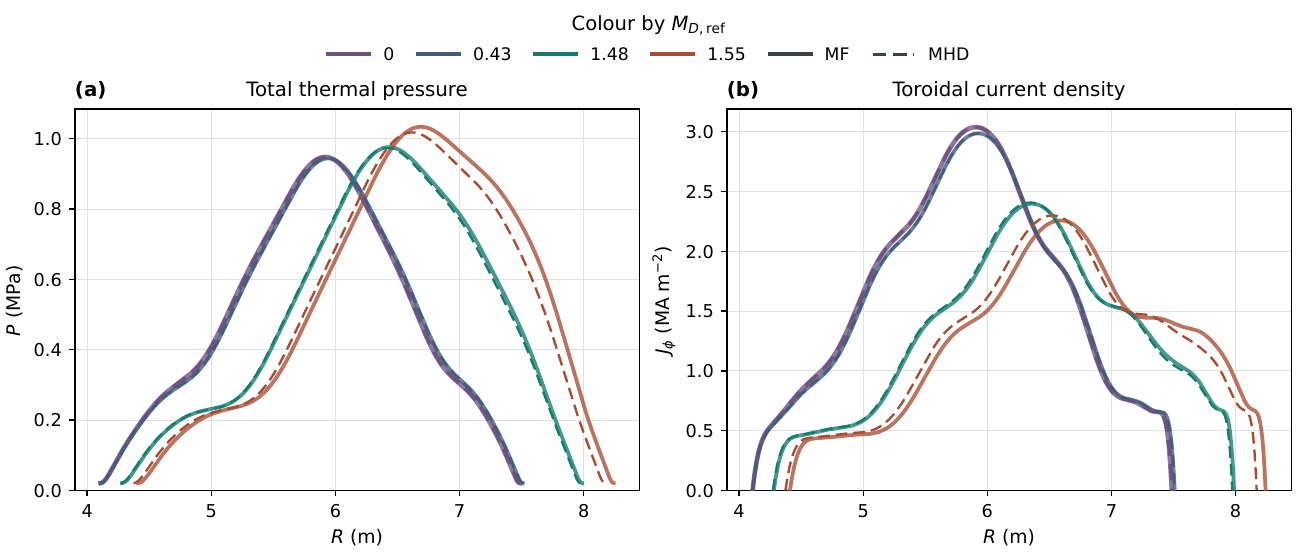}
\caption{Total thermal pressure (a) and toroidal current density (b) in pure DT at \(Z=0.50\ \mathrm m\). Colour denotes the deuterium reference thermal Mach number. Solid and dashed lines denote MF and rotating MHD, respectively. The zero-rotation profiles overlap.}
\label{fig:4-3}
\end{figure}

At \(240\) and \(252\ \mathrm{krad\,s^{-1}}\), the total-pressure and toroidal-current profiles shift towards larger \(R\) relative to zero rotation, with a larger difference between MF and rotating MHD at \(252\ \mathrm{krad\,s^{-1}}\) (Fig.~\ref{fig:4-3}). Section~\ref{sec:4-3} evaluates the poloidal-flux contributions from the MF--MHD differences in the pressure-derivative and \(FF'\) current terms and uses them to estimate the outboard LCFS crossing displacement.

To distinguish the boundary displacement caused by rotation from the additional change between the multi-fluid and rotating MHD equilibria, we examine pure-DT cases at \(11\ \mathrm{MA}\) with an on-axis angular velocity of \(70\ \mathrm{krad\,s^{-1}}\).

Both rotating profiles use the same lower-single-null configuration. We compare each with the zero-rotation MF equilibrium of that configuration, holding the coil currents, reference densities and temperatures, total current and all other prescribed inputs fixed.
\par
\begin{samepage}
\noindent They use the same \(F_{\mathrm{vac}},\alpha_F,\beta_F\) as the \(15\ \mathrm{MA}\) cases, with \(F^2-F_{\mathrm{vac}}^2=c(F_b^2-F_{\mathrm{vac}}^2)\), where \(c=0.53778\) and \(F_b\) is given by Eq.~\eqref{eq:2-5}. The total-current constraint determines \(\gamma\). The uniform exterior pressure is \(10.54\ \mathrm{kPa}\). The common ion temperature is \(23.0\ \mathrm{keV}\) on axis and \(9.90\ \mathrm{keV}\) at \(x=0.50\).
\par
\end{samepage}

\begin{figure}[!htbp]
\centering
\includegraphics[width=\linewidth]{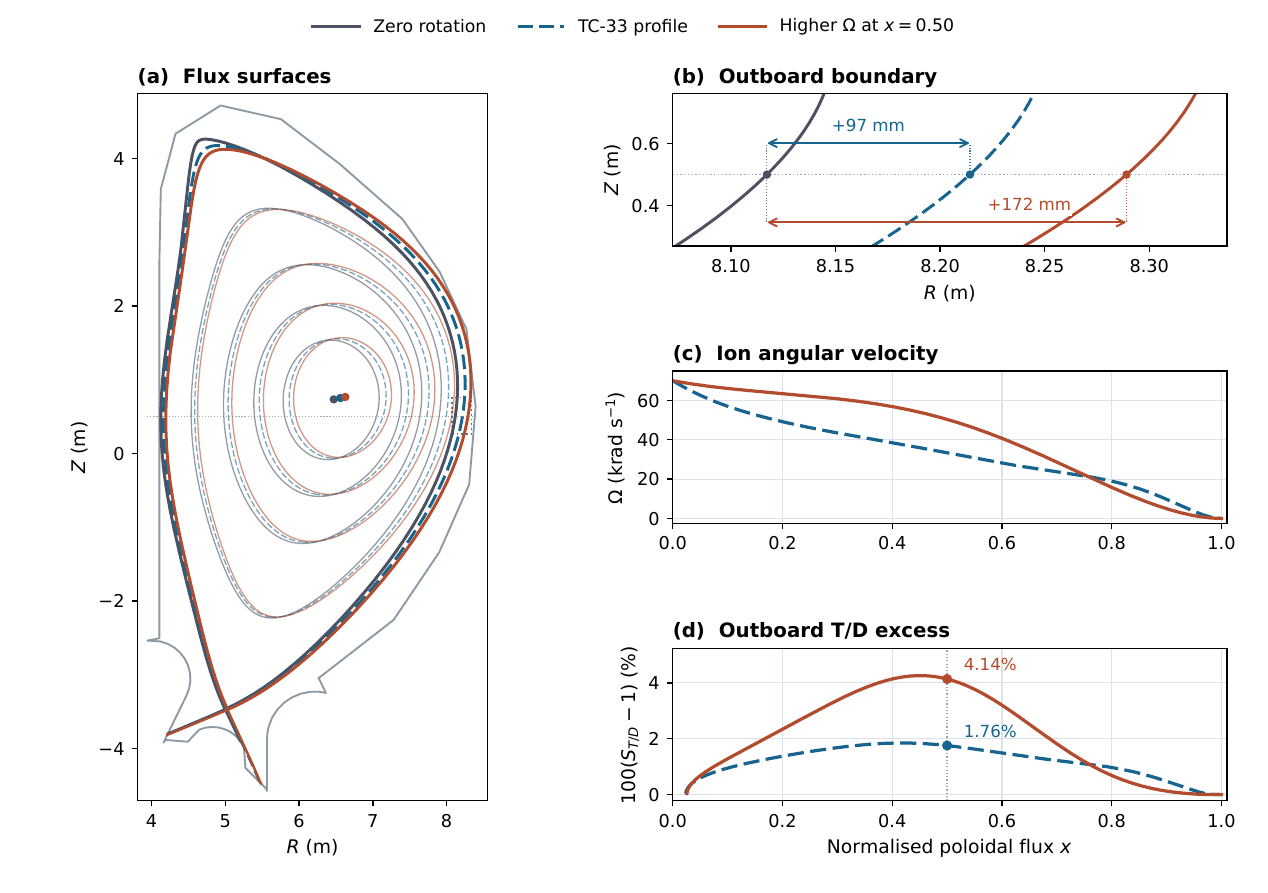}
\caption{\(11\ \mathrm{MA}\) pure-DT MF equilibria. (a) Flux surfaces and separatrices. (b) Outboard boundary near \(Z=0.50\ \mathrm m\), with radial displacements from the zero-rotation MF equilibrium for this configuration. (c) Ion angular velocity. (d) Percentage increase in the outboard T/D ratio relative to the inboard ratio on the same surface at \(Z=0.50\ \mathrm m\).}
\label{fig:4-11ma}
\end{figure}

The first rotating profile has a radial shape based on the TC-33 reference case~\cite{bourdelle2025tc33}, scaled to \(70\ \mathrm{krad\,s^{-1}}\) on axis. The second combines 30\% of that normalised shape with 70\% of \((1-x^4)^3\) and uses the same axis value.

On the \(Z=0.50\ \mathrm m\) cut, the outboard MF boundary crossing moves outward by \(97\ \mathrm{mm}\) for the TC-33 profile and \(172\ \mathrm{mm}\) for the profile with the higher angular velocity at \(x=0.50\), relative to the zero-rotation MF equilibrium with the other inputs fixed [Fig.~\ref{fig:4-11ma}(a,b)]. Both rotating equilibria retain the lower single-null configuration.

For either prescribed angular-velocity profile, the rotating-MHD and MF outboard crossings differ by less than \(0.2\ \mathrm{mm}\), and both full LCFS RMS distances are below \(0.2\ \mathrm{mm}\). At \(x=0.50\) the second profile raises the angular velocity from \(33.3\) to \(50.4\ \mathrm{krad\,s^{-1}}\), while at \(x=0.80\) it lowers it from \(19.0\) to \(15.8\ \mathrm{krad\,s^{-1}}\). On the \(x=0.50\) surface, MF gives outboard T/D density ratios \(1.76\%\) and \(4.14\%\) above the corresponding inboard values [Fig.~\ref{fig:4-11ma}(c,d)]. In these cases, rotating MHD captures the boundary displacement caused by rotation, while the multi-fluid solution resolves the accompanying D--T separation.

\FloatBarrier
\subsection{Effect of ion composition}
\label{sec:4-2}

We use the \(15\ \mathrm{MA}\) pure-DT equilibrium at \(252\ \mathrm{krad\,s^{-1}}\) as the control and set the reference helium fraction \(f_{\mathrm{He}}=N_{\mathrm{He}}/N_e\) to \(2.5\%\) and \(5\%\) in the helium cases. These two fractions lie in the few-percent range measured in helium-injection experiments on ASDEX Upgrade and JT-60U~\cite{herschel2026gas,sakasai1999helium}. Transport modelling for a \(15\ \mathrm{MA}\), \(5.3\ \mathrm T\) ITER DT scenario gives a helium-to-electron density ratio of \(4\)--\(5\%\) on axis~\cite{garzotti2019integrated}.

For a prescribed \(N_{\mathrm{He}}(x)\), we determine \(N_D(x)\) and \(N_T(x)\) at each flux label using quasineutrality \(N_D+N_T+2N_{\mathrm{He}}=N_e\) and the mass-density constraint \(m_DN_D+m_TN_T+m_{\mathrm{He}}N_{\mathrm{He}}=\rho_0^{\mathrm{DT}}(x)\).

\pagebreak
The composition scan uses the prescribed baseline ratio \(\tau_0(x)=\theta_i(x)/\theta_e(x)\), i.e., \(k=1\) in Eq.~\eqref{eq:4-4}. At fixed \(P_0(x)\) and \(N_e(x)\), inserting the total reference ion density \((1-f_{\mathrm{He}})N_e\) from quasineutrality into Eq.~\eqref{eq:2-6} gives
\[
\theta_i(x)=
\frac{\tau_0(x)P_0(x)}
{N_e(x)\left[1+(1-f_{\mathrm{He}})\tau_0(x)\right]}.
\]
Increasing \(f_{\mathrm{He}}\) at this fixed temperature ratio raises both the ion and electron temperatures (Table~\ref{tab:4-he-inputs}).

\begin{figure}[!tp]
\centering
\captionsetup{skip=5pt}
\includegraphics[width=\linewidth,height=.735\textheight,keepaspectratio]{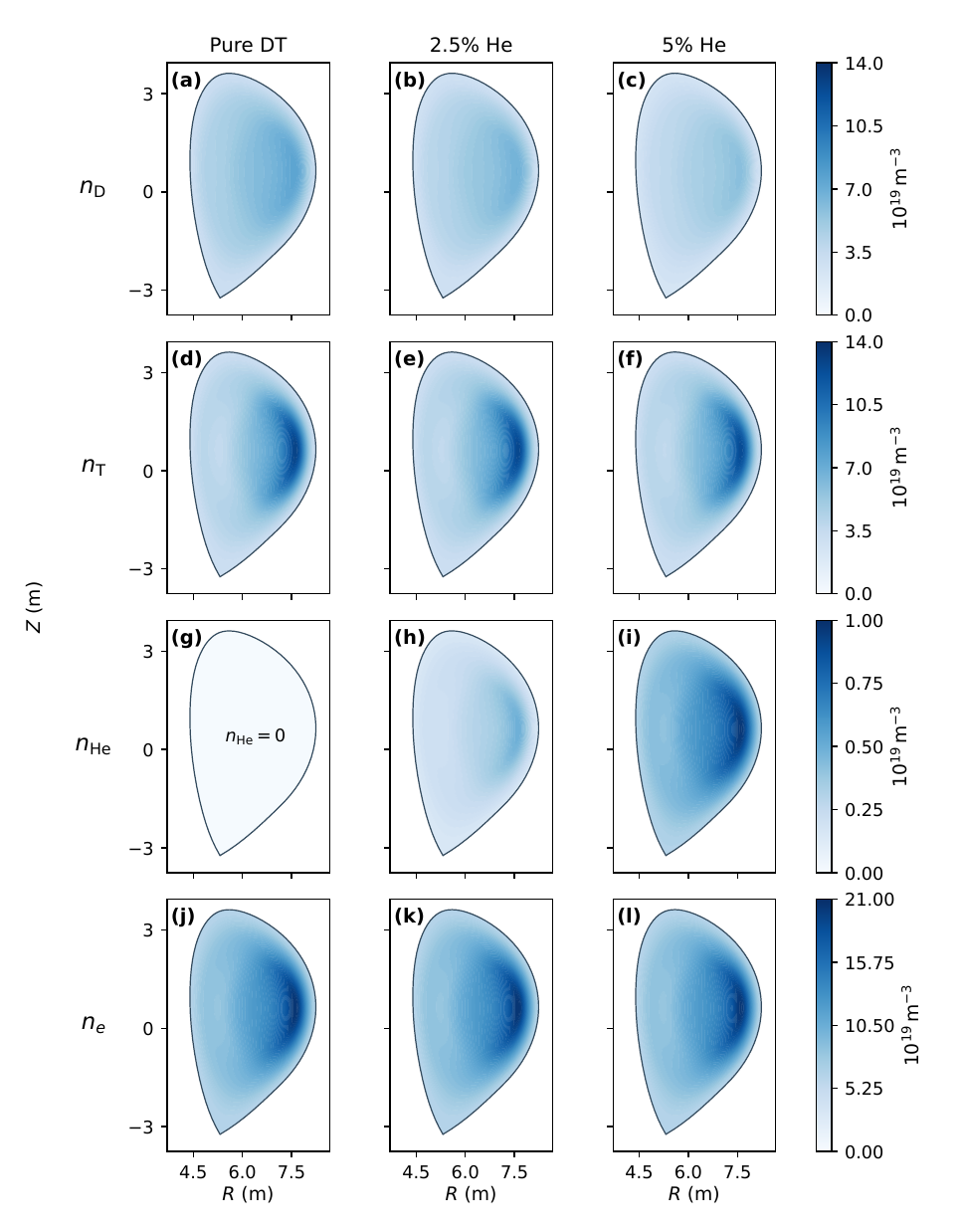}
\caption{Absolute species densities for pure DT, \(2.5\%\) He and \(5\%\) He (columns). The rows show D, T, He and electrons. Each species uses one scale across the three cases, and D and T share a scale.}
\label{fig:4-4}
\end{figure}

\begin{table}[H]
\centering
\small
\captionsetup{skip=5pt}
\caption{Reference composition and on-axis temperatures for the helium scan. The cases share the \(N_e(x)\), \(\rho_0(x)\), \(P_0(x)\) and baseline \(T_i/T_e\) profiles.}
\label{tab:4-he-inputs}
\begin{tabular}{@{}rrrrr@{}}
\toprule
\(N_{\mathrm{He}}/N_e\) (\%) & \(N_D/N_e\) & \(N_T/N_e\) & \(T_i(0)\) (keV) & \(T_e(0)\) (keV) \\
\midrule
0 & 0.500 & 0.500 & 21.1 & 28.5 \\
2.5 & 0.449 & 0.501 & 21.3 & 28.8 \\
5 & 0.399 & 0.501 & 21.5 & 29.1 \\
\bottomrule
\end{tabular}
\end{table}
Adding helium raises the T/D density ratio at both sampled sides of the \(x=0.50\) surface while slightly reducing the inboard--outboard contrast. Figure~\ref{fig:4-4} gives the corresponding species densities. On that surface at \(Z=0.50\ \mathrm m\), \(S_{T/D}\) decreases from \(2.13\) through \(2.11\) to \(2.09\) for \(0\), \(2.5\%\) and \(5\%\) He, while the reference ratio \(N_T/N_D\) rises from \(1\) through \(1.11\) to \(1.26\). Higher local T/D ratios thus accompany weaker centrifugal separation at these crossings.

Because D and T are singly charged and share \(\theta_i\) and \(\Omega\) in DT--He, Eq.~\eqref{eq:4-3} and the resulting expression for \(S_{T/D}\) also apply. The reference ratio \(N_T/N_D\) scales the local density ratio but cancels from \(S_{T/D}\).

The rise in \(T_i\) gives the larger reduction in \(S_{T/D}\) when the temperature and geometry changes are substituted separately into the separation-factor relation in Section~\ref{sec:4-1}. From pure DT to \(5\%\) He on this surface, \(T_i\) rises from \(8.07\) to \(8.25\ \mathrm{keV}\), and \(R_{\mathrm{out}}^2-R_{\mathrm{in}}^2\) decreases from \(31.60\) to \(31.47\ \mathrm{m^2}\).

\begin{figure}[!tp]
\centering
\captionsetup{skip=5pt}
\includegraphics[width=\linewidth]{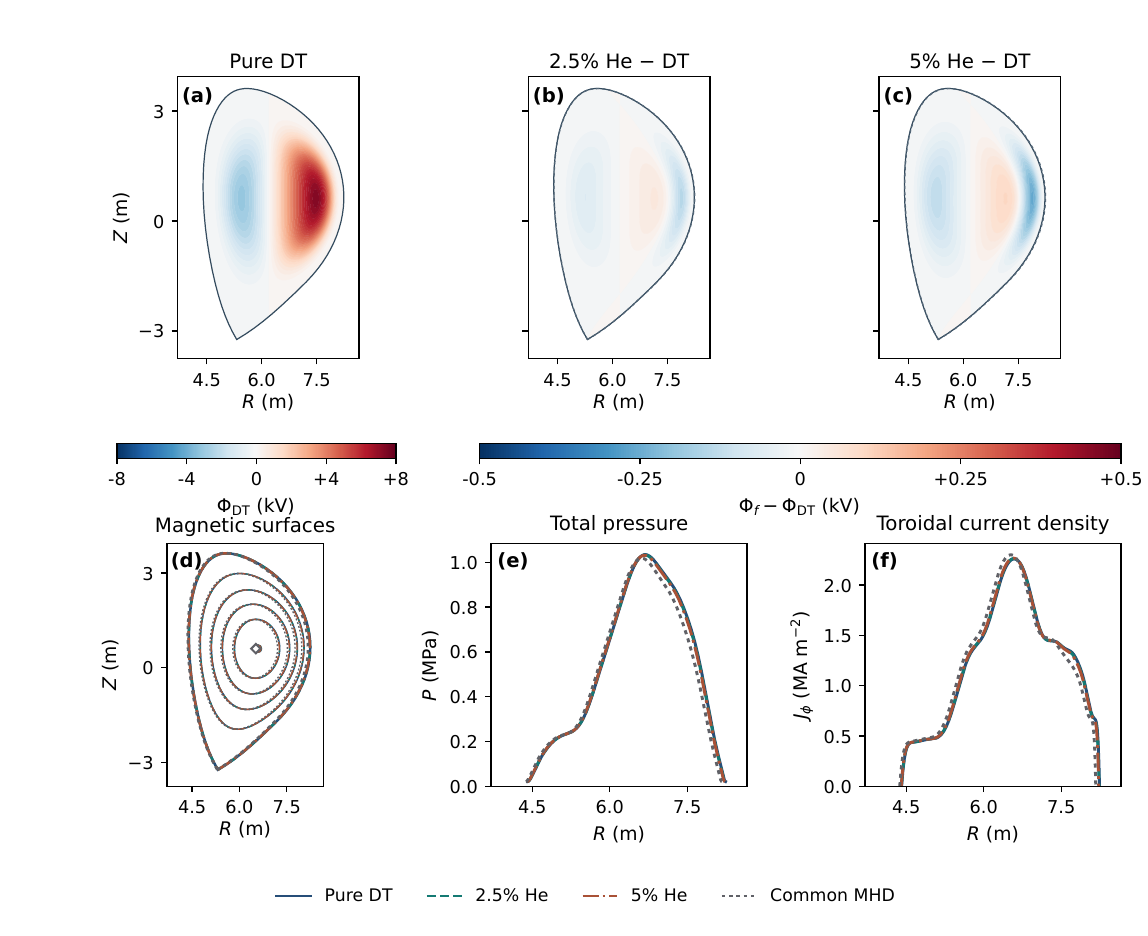}
\caption{Potential in pure DT (a) and differences from pure DT for \(2.5\%\) and \(5\%\) He (b,c). Each difference is shown in the overlap of the regions enclosed by the LCFSs of the pure-DT and corresponding helium equilibria. Solid and dashed curves mark the helium and pure-DT LCFS, respectively. Panels (d--f) compare magnetic surfaces and the total pressure and toroidal current density at \(Z=0.50\ \mathrm m\) with the common rotating-MHD result.}
\label{fig:4-5}
\end{figure}

Adding helium produces both positive and negative potential differences on the low-field side, with a negative region near the outer edge (Fig.~\ref{fig:4-5}(b,c)). Each map subtracts the pure-DT potential at the same \((R,Z)\), using the common reference \(\Phi(R_0,x)=0\). That spatial point can lie on different flux surfaces in the two equilibria.

The outboard crossing of the MF \(x=0.50\) surface moves inward as the reference helium fraction rises. On the \(Z=0.50\ \mathrm m\) cut, the fine-grid crossing shifts inward by \(8.4\ \mathrm{mm}\) at \(2.5\%\) He and by \(15.5\ \mathrm{mm}\) at \(5\%\) He relative to pure DT. Both grids give these inward shifts, with magnitudes changing by about \(3\)--\(4\ \mathrm{mm}\) on refinement (Table~\ref{tab:3-3}). At \(5\%\) He the total-pressure and toroidal-current-density peaks also move inward, by about \(14\) and \(13\ \mathrm{mm}\), respectively [Fig.~\ref{fig:4-5}(d--f)].

\FloatBarrier
\subsection{Effect of the ion-to-electron temperature ratio}
\label{sec:4-3}

For pure DT and DT with \(5\%\) He, we vary the ion-to-electron temperature ratio at fixed \(P_0(x)\), \(\Omega(x)\) and species reference densities for each composition. We use three multipliers \(k\) of the baseline temperature-ratio profile \(\tau_0(x)\)
\[
\frac{T_i(x)}{T_e(x)}=k\,\tau_0(x),\qquad
\tau_0(x)=0.795-0.055(1-x)^2,\qquad k=0.50,\ 1.00,\ 1.20 .
\tag{4.4}
\label{eq:4-4}
\]
Substituting \(T_i(x)=k\tau_0(x)T_e(x)\) into the pressure sum in Eq.~\eqref{eq:2-6} gives \(T_e(x)\), and the prescribed ratio gives the common ion temperature \(T_i(x)\). For pure DT with equal D and T reference densities, fixed \(P_0\) and \(N_e\) give \(\theta_i+\theta_e=P_0/N_e\). Lowering \(T_i\) increases the magnitude of the centrifugal energy difference relative to the ion thermal energy. At the same \(R\), \(x\) and \(\Omega\), it raises \(P_{\mathrm{MF}}/P_{\mathrm{MHD}}\) wherever \(\chi\ne0\), by the exact relation in Section~\ref{sec:4-1}. Each equilibrium determines the particle inventories and thermal energy.

A lower \(k\) raises each ion reference thermal Mach number because \(T_i\) decreases at the same angular velocity (Table~\ref{tab:4-temperature}). The three values of \(k\) give on-axis ratios \(0.370\), \(0.740\) and \(0.888\), with \(T_e>T_i\) throughout each case.

\begin{table}[H]
\centering
\captionsetup{skip=5pt}
\caption{Temperature ratios, reference thermal Mach numbers, T/D separation and geometric changes.}
\label{tab:4-temperature}
\small
\begin{tabular}{@{}lrrrrr@{}}
\toprule
Mixture & \(k\) & \((T_i/T_e)_{x=0}\) & \(M_{D,\mathrm{ref}}\) & \(M_{T,\mathrm{ref}}\) & \(M_{\mathrm{He},\mathrm{ref}}\) \\
\midrule
DT & 0.50 & 0.370 & 1.95 & 2.39 & \textemdash \\
 & 1.00 & 0.740 & 1.55 & 1.90 & \textemdash \\
 & 1.20 & 0.888 & 1.48 & 1.81 & \textemdash \\
DT--He (5\%) & 0.50 & 0.370 & 1.94 & 2.37 & 2.73 \\
 & 1.00 & 0.740 & 1.54 & 1.88 & 2.17 \\
 & 1.20 & 0.888 & 1.46 & 1.79 & 2.06 \\
\bottomrule
\end{tabular}\par\smallskip
\setlength{\tabcolsep}{7pt}
\begin{tabular}{@{}lrrrrrr@{}}
\toprule
Mixture & \(k\) & \(S_{\mathrm{MF}}\) & \(S_{\mathrm{post}}\) & \shortstack{Relative\\difference (\%)} &
\(\Delta R_{\mathrm{LFS}}^{\mathrm{MF}}\) (mm) & \(d_{\mathrm{LCFS}}\) (mm) \\
\midrule
DT & 0.50 & 3.34 & 3.18 & \(-4.70\) & \(+75\) & 84 \\
 & 1.00 & 2.13 & 2.10 & \(-1.54\) & 0 & 42 \\
 & 1.20 & 1.98 & 1.96 & \(-1.20\) & \(-11\) & 36 \\
DT--He (5\%) & 0.50 & 3.25 & 3.13 & \(-3.87\) & \(+65\) & 70 \\
 & 1.00 & 2.09 & 2.06 & \(-1.20\) & 0 & 34 \\
 & 1.20 & 1.94 & 1.93 & \(-0.98\) & \(-6\) & 30 \\
\bottomrule
\end{tabular}
\par\smallskip{\footnotesize\raggedright Fine-grid results. The T/D separation factors \(S_{\mathrm{MF}}\) and \(S_{\mathrm{post}}\) use the inboard and outboard crossings at \(Z=0.50\ \mathrm m\) of the respective MF and common rotating MHD \(x=0.50\) surfaces. The relative difference in percent is \(100(S_{\mathrm{post}}/S_{\mathrm{MF}}-1)\), calculated before rounding. A negative value indicates a lower MHD-based estimate. At the same \(x,Z\), \(\Delta R_{\mathrm{LFS}}^{\mathrm{MF}}\) is the MF outboard shift from \(k=1\) at the same composition, positive outward. \(d_{\mathrm{LCFS}}\) is the full-contour symmetric RMS distance to the common rotating MHD LCFS. The four \(x=0.50\) crossing shifts in Table~\ref{tab:3-3} change by \(3\)--\(4\ \mathrm{mm}\) on refinement.\par}
\end{table}

\begin{figure}[H]
\centering
\includegraphics[width=\linewidth]{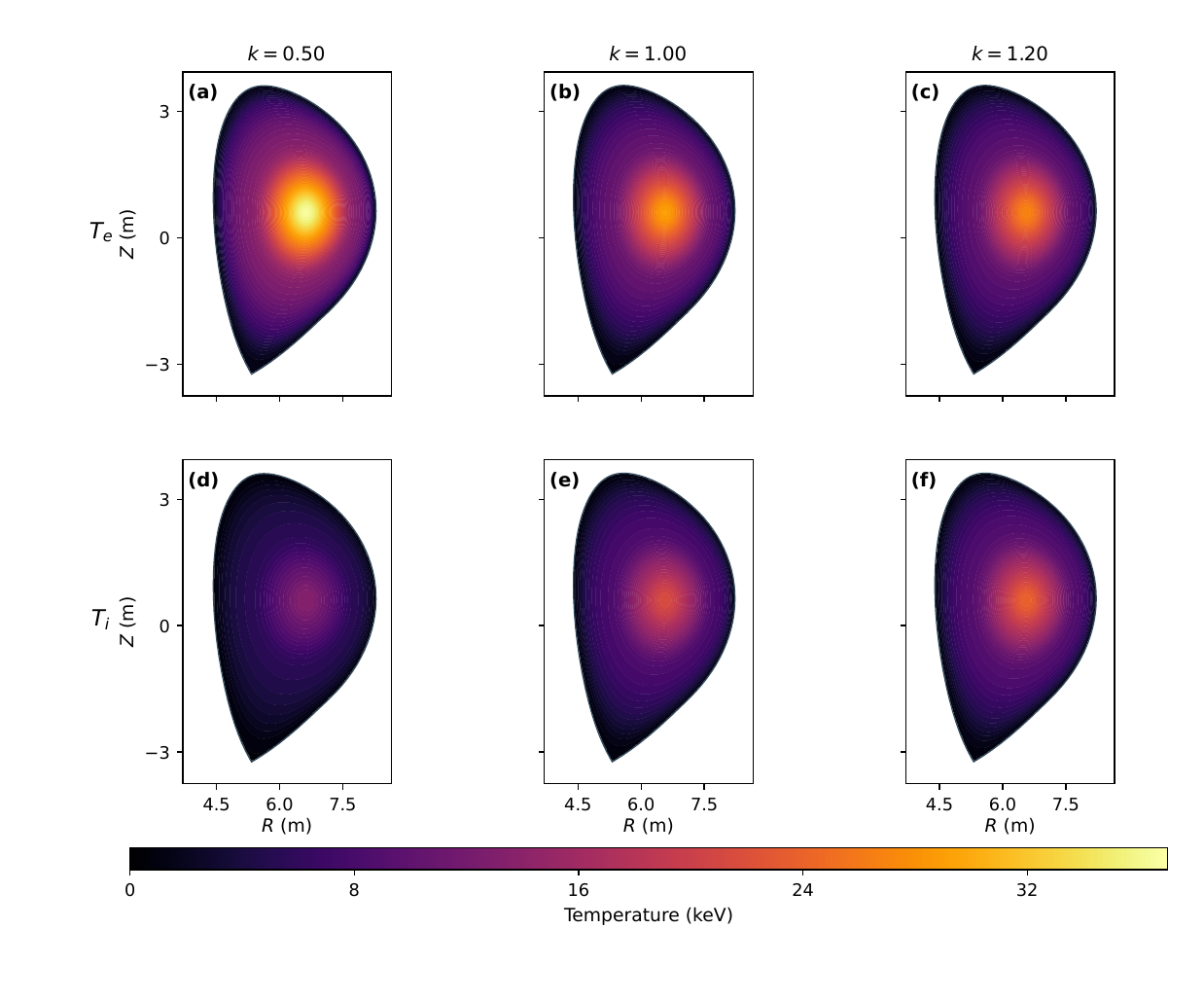}
\caption{Two-dimensional electron (top) and common ion (bottom) temperatures for \(5\%\) He at \(252\ \mathrm{krad\,s^{-1}}\). The columns give the three values of \(k\), and all panels share a temperature scale in keV.}
\label{fig:4-6}
\end{figure}

At fixed reference thermal pressure, decreasing \(k\) raises the electron temperature and lowers the common ion temperature. In the \(5\%\) He cases, decreasing \(k\) from \(1.20\) to \(0.50\) raises the on-axis \(T_e\) from \(26.9\) to \(36.7\ \mathrm{keV}\) and lowers the on-axis \(T_i\) from \(23.9\) to \(13.6\ \mathrm{keV}\). Figure~\ref{fig:4-6} shows the prescribed temperatures over the self-consistent \(5\%\) He magnetic surfaces.

\begin{figure}[H]
\centering
\includegraphics[width=\linewidth]{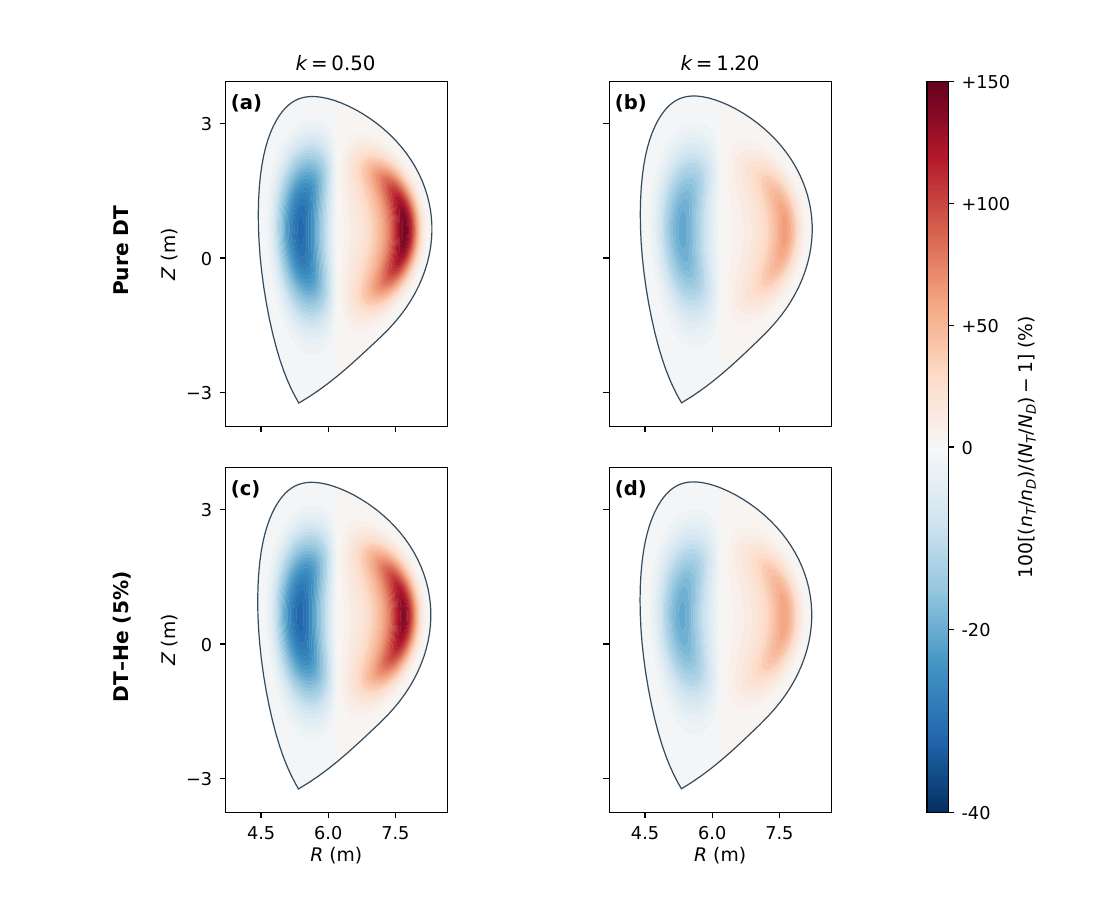}
\caption{Percentage change in the local T/D density ratio relative to its reference value on the same surface. Rows show pure DT and \(5\%\) He. Columns show \(k=0.50\) and \(1.20\), with a common signed scale. The colour scale is centred at zero, with separate linear ranges of \(-40\%\) to 0 and 0 to \(+150\%\).}
\label{fig:4-7}
\end{figure}

\begin{samepage}
To measure local tritium enrichment or depletion relative to the reference ratio on each surface, we use the percentage change
\[
E_{T/D}(R,Z)=100\!\left[\frac{n_T(R,Z)/n_D(R,Z)}{N_T(x)/N_D(x)}-1\right].
\tag{4.5}
\label{eq:4-5}
\]
Here \(E_{T/D}\) is in percent, with \(N_T/N_D=1\) for pure DT and about \(1.26\) for \(5\%\) He (Fig.~\ref{fig:4-7}).
\par
\end{samepage}

Both compositions show a stronger inboard--outboard contrast in the local T/D ratio at \(k=0.50\) than at \(k=1.20\) (Fig.~\ref{fig:4-7}). On the \(x=0.50\) surface, the separation factor increases from \(1.98\) to \(3.34\) in pure DT and from \(1.94\) to \(3.25\) with \(5\%\) He (Table~\ref{tab:4-temperature}).

Across the six temperature-ratio cases, the T/D separation factors estimated on the common rotating MHD surface are \(0.98\%\)--\(4.70\%\) below the corresponding self-consistent multi-fluid values (Table~\ref{tab:4-temperature}). We evaluate the separation-factor relation derived from Eq.~\eqref{eq:4-3} using the inboard and outboard radii of the MHD \(x=0.50\) surface at \(Z=0.50\ \mathrm m\), the ion temperature \(T_i(0.50)\) for each case and the common angular velocity \(\Omega(0.50)\). The multi-fluid values use the crossings of each equilibrium's own surface at the same flux label and height. For pure DT at \(k=0.50\), the estimate is \(3.18\) and the self-consistent value is \(3.34\).

\begin{figure}[H]
\centering
\includegraphics[width=\linewidth,height=.600\textheight,keepaspectratio]{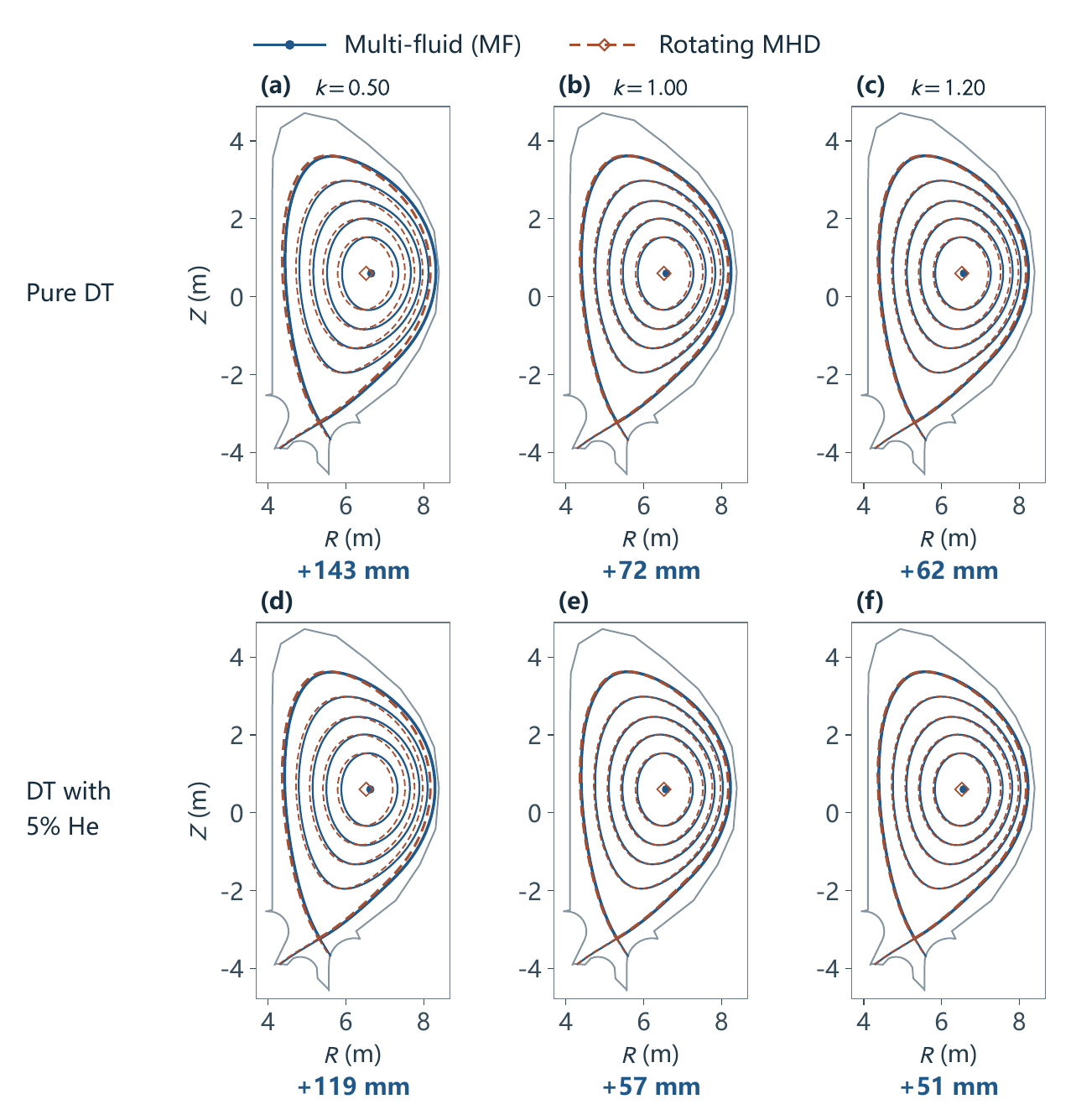}
\caption{Multi-fluid and common rotating MHD surfaces for pure DT (top) and DT with \(5\%\) He (bottom) at three values of \(k\) in \(T_i/T_e=k\tau_0(x)\). Each label gives the radial displacement of the outboard LCFS crossing, MF minus MHD, at \(Z=0.50\ \mathrm m\).}
\label{fig:4-8}
\end{figure}

In pure DT, lowering \(k\) from \(1.20\) to \(0.50\) increases the full-contour LCFS RMS distance from the common rotating-MHD reference from \(36\) to \(84\ \mathrm{mm}\) (Table~\ref{tab:4-temperature}). At \(k=0.50\), the outboard MF LCFS crossing at \(Z=0.50\ \mathrm m\) lies \(143\ \mathrm{mm}\) farther out than the rotating MHD crossing. This local difference changes by \(0.566\ \mathrm{mm}\) upon grid refinement (Table~\ref{tab:3-3}). The \(5\%\) He cases show the same trend (Fig.~\ref{fig:4-8}).

Close separation estimates therefore coexist with different LCFS positions. For prescribed species inputs, \(S_{T/D}\) depends on geometry only through \(R_{\mathrm{out}}^2-R_{\mathrm{in}}^2\) at the two \(x=0.50\) crossings. This combination leaves the full shape of that surface and the LCFS undetermined.

\begin{samepage}
The \(15\ \mathrm{MA}\) composition and temperature-ratio scans at \(252\ \mathrm{krad\,s^{-1}}\) also retain the same LCFS thermal pressure, since \(\Omega(1)=0\) and \(P_0(1)\) is unchanged. Their different boundaries therefore occur with the same edge thermal pressure, coil currents and total plasma current. We next calculate the flux changes produced by the pressure-derivative and \(FF'\) current terms and use them to estimate the displacement of the outboard LCFS crossing.
\par
\end{samepage}

We use four \(15\ \mathrm{MA}\) pure-DT comparisons. At the baseline temperature ratio (\(k=1.00\)), we compare MF and rotating MHD at \(240\), \(246\) and \(252\ \mathrm{krad\,s^{-1}}\). For the \(k=0.50\) MF case at \(252\ \mathrm{krad\,s^{-1}}\), we use the same rotating MHD equilibrium as for \(k=1.00\) at that angular velocity. In each pair, \(X\) and \(B\) denote the rotating MHD equilibrium's lower X point and outboard LCFS crossing on the \(Z=0.50\ \mathrm m\) cut.

We use the pressure-derivative and toroidal-field-function terms \(J_P\) and \(J_F\) of Eq.~\eqref{eq:2-11}. On the common grid, \(\Delta J_{\alpha,j}=J_{\alpha,j}^{\mathrm{MF}}-J_{\alpha,j}^{\mathrm{MHD}}\) for \(\alpha=P,F\), with each equilibrium's own plasma current distribution.

\begin{samepage}
The Green function \(G\) gives the poloidal flux per unit toroidal ring current. With grid spacings \(h_R,h_Z\), the sum over the whole computational cross-section gives the contribution from each current term to the change in \(\psi(X)-\psi(B)\),
\[
A_\alpha=\sum_j [G(X,j)-G(B,j)]\,\Delta J_{\alpha,j}\,h_Rh_Z,\qquad \alpha=P,F.
\]
\par
\end{samepage}
\begin{samepage}
The coil flux is the same in each pair. The correction \(\eta_X=\psi_{b,\mathrm{MF}}-\psi_{\mathrm{MF}}(X)\) accounts for the shift of the X point between the two equilibria. Expanding the MF flux radially from \(B\), the first-order estimate of the MF-minus-MHD crossing displacement is
\[
\widehat{\Delta R}_{\mathrm{out}}
=\frac{A_P+A_F+\eta_X}{(\partial_R\psi_{\mathrm{MF}})_B},
\]
where the denominator is the radial gradient of the converged MF total flux at \(B\). Positive displacement means the MF crossing is farther out.
\par
\end{samepage}

\begin{table}[!htbp]
\centering
\captionsetup{skip=5pt}
\caption{Current contributions to the flux difference and outboard LCFS shift at \(Z=0.50\ \mathrm m\) for \(15\ \mathrm{MA}\) pure DT on the \(257\times513\) grid. \(A_P\) and \(A_F\) are the current contributions to \(\psi(X)-\psi(B)\), and \(\eta_X\) corrects for X-point motion. \(X\) and \(B\) are the lower X point and outboard LCFS crossing of the rotating MHD reference. The gradient is of the converged MF total flux at \(B\). Shifts are MF minus rotating MHD, positive outward.}
\label{tab:4-current-flux}
\small
\setlength{\tabcolsep}{3.8pt}
\begin{tabular}{@{}rrrrrrrr@{}}
\toprule
\shortstack{\(\Omega_0\)\\\((\mathrm{krad\,s^{-1}})\)} & \(k\) &
\shortstack{\(A_P\)\\(\(\mathrm{Wb\,rad^{-1}}\))} & \shortstack{\(A_F\)\\(\(\mathrm{Wb\,rad^{-1}}\))} &
\shortstack{\(\eta_X\)\\(\(\mathrm{Wb\,rad^{-1}}\))} &
\shortstack{\((\partial_R\psi_{\mathrm{MF}})_B\)\\(\(\mathrm{Wb\,rad^{-1}\,m^{-1}}\))} &
\shortstack{First-order\\(mm)} & \shortstack{Direct\\(mm)} \\
\midrule
240 & 1.00 & \(-0.182\) & \(-0.0116\) & \(4.15\times10^{-5}\) & \(-12.7\) & \(+15.16\) & \(+15.17\) \\
246 & 1.00 & \(-0.337\) & \(-0.0220\) & \(1.39\times10^{-4}\) & \(-13.0\) & \(+27.65\) & \(+27.65\) \\
252 & 1.00 & \(-0.917\) & \(-0.0422\) & \(9.53\times10^{-4}\) & \(-13.2\) & \(+72.34\) & \(+71.87\) \\
252 & 0.50 & \(-1.85\) & \(-0.0709\) & \(3.74\times10^{-3}\) & \(-13.2\) & \(+145.1\) & \(+142.6\) \\
\bottomrule
\end{tabular}
\end{table}

The pressure-derivative term accounts for about \(94\%\)--\(96\%\) of the current-induced change in \(\psi(X)-\psi(B)\) in the four comparisons (Table~\ref{tab:4-current-flux}), as measured by \(A_P/(A_P+A_F)\). Equal total plasma currents give a zero cross-sectional integral of the total toroidal current-density difference, whereas the spatially varying weight \(G(X,j)-G(B,j)\) allows current redistribution to change the flux difference and boundary position. At \(k=1.00\), increasing \(\Omega_0\) from \(240\) to \(252\ \mathrm{krad\,s^{-1}}\) increases the magnitude of \(A_P+A_F+\eta_X\) by a factor of \(4.96\), while the magnitude of \((\partial_R\psi_{\mathrm{MF}})_B\) increases by about \(4\%\). After accounting for X-point motion, the estimated shifts agree with direct tracing to within \(2.5\ \mathrm{mm}\). For pure DT at \(k=0.50\), they are \(145.1\) and \(142.6\ \mathrm{mm}\), respectively.

\section{Discussion and conclusions}
\label{sec:5}

MFEQ-FB extends the reduced rotating multi-fluid model to a self-consistent calculation of species distributions and the plasma boundary. Under common ion rotation, analytical comparison at the same \(R\), \(x\) and reference profiles gives \(P_{\mathrm{MF}}\geq P_{\mathrm{MHD}}\). At fixed \(R\), \(x\), reference densities and temperatures, the pressure ratio is nondecreasing in \(|\Omega|\). For \(\Omega\ne0\) and \(R\ne R_0\), the pressures are equal exactly when all species present have the same relative density \(n_s/N_s\) (Appendix~\ref{app:pressure-ratio}).

At an on-axis angular velocity of \(70\ \mathrm{krad\,s^{-1}}\), the \(15\ \mathrm{MA}\) pure-DT MF and rotating MHD boundaries nearly coincide, while the outboard T/D density ratio exceeds the inboard value by \(4.8\%\) on the \(x=0.50\) surface at \(Z=0.50\ \mathrm m\). This shows that D--T separation can occur on a magnetic surface even when the multi-fluid and rotating MHD boundaries nearly coincide. At fixed geometry, the weak-rotation expansion gives separation at order \(\lambda^2\) and a pressure difference at order \(\lambda^4\) (Section~\ref{sec:4-1}). In the separate \(11\ \mathrm{MA}\) configuration at the same on-axis angular velocity, rotating MHD captures the appreciable boundary shifts that both prescribed profiles produce relative to that configuration's zero-rotation MF equilibrium.

Changing the ion composition or the ion-to-electron temperature ratio produces distinct MF boundaries under a common rotating MHD reference. The composition and temperature-ratio scans at \(15\ \mathrm{MA}\) and \(\Omega_0=252\ \mathrm{krad\,s^{-1}}\) hold the reference profiles \(N_e,\rho_0,P_0,\Omega\), coil currents and total plasma current fixed. In pure DT, lowering the temperature-ratio profile so that its on-axis value decreases from \(0.888\) to \(0.370\) increases the RMS distance between the MF and MHD LCFSs from \(36\) to \(84\ \mathrm{mm}\).

Across the six temperature-ratio cases, T/D separation factors estimated on the common MHD geometry at \(x=0.50\) and \(Z=0.50\ \mathrm m\) remain close to the self-consistent MF values (Table~\ref{tab:4-temperature}). At the lowest pure-DT temperature ratio, the estimate is \(3.18\) and the MF value is \(3.34\), while the MF outboard LCFS crossing on the same cut lies about \(143\ \mathrm{mm}\) beyond the MHD crossing. Using prescribed species inputs on MHD surfaces estimates T/D separation; including the species pressures in the equilibrium calculation determines the self-consistent geometry.

The pressure-derivative contribution accounts for most of the current-induced change in poloidal flux difference between the reference X point and the outboard LCFS crossing in the four pure-DT comparisons (Table~\ref{tab:4-current-flux}). The total current is unchanged, but its spatial distribution differs. Including X-point motion, first-order estimates of the outboard LCFS crossing shifts agree with direct tracing to within \(2.5\ \mathrm{mm}\). A next application is to prescribe a common target X-point position and outboard LCFS radius at a specified height, and compare the poloidal-field coil currents required by the two models.

\section*{Acknowledgments}
This research is supported by the National MCF Energy R\&D Program of China (Grant Nos. 2024YFE03000102 and 2022YFE03090000) and the National Natural Science Foundation of China (Grant Nos. 12435014 and 12475214).

ChatGPT (OpenAI) was used to assist with language editing and mathematical analysis. The authors reviewed and verified the outputs and take full responsibility for the manuscript.

\clearpage
\appendix
\makeatletter
\renewcommand{\@seccntformat}[1]{\appendixname~\csname the#1\endcsname.\enspace}
\makeatother
\numberwithin{equation}{section}

\section{Assessing the use of poloidal flux to label species profiles}
\label{app:ion-label}

We assess the size of the drift-surface correction to the poloidal-flux labels of the prescribed profiles in Section~\ref{sec:2-1} for D, T and He ions sharing the angular velocity \(\Omega_i(x)\).

The drift-surface label in Section~2.7 and Appendix~B of Xie et al.~\cite{xie2026reduced} gives the normalised offset
\[
\delta x_s=\frac{Y_s-\psi}{\Delta\psi}
=\frac{m_sR^2\Omega_i(x)}{Z_se\,\Delta\psi},
\qquad \Delta\psi=\psi_b-\psi_a .
\]
Each equilibrium uses its own signed flux difference \(\Delta\psi\). For a prescribed input profile \(Q\in\{N_s,T_s,\Omega_i\}\), the product \(\delta x_s Q_x/Q\) measures the first-order relative change in that input from shifting its label. For \(\Omega_i\), we evaluate the product as \(m_sR^2\Omega_{i,x}/(Z_se\,\Delta\psi)\), which also gives its limit at \(\Omega_i=0\).

The four column maxima for each ion can occur at different nodes.

\begin{table}[H]
\centering
\small
\setlength{\tabcolsep}{6pt}
\captionsetup{skip=5pt}
\caption{Offsets of ion drift-surface labels from the poloidal flux and the resulting first-order estimates of relative changes in the prescribed profiles for the ten \(15\ \mathrm{MA}\) multi-fluid equilibria. The four diagnostics are reported as maximum absolute values, expressed as percentages, over the saved grid nodes within the LCFS of each equilibrium on the \(257\times513\) grid. The offset uses each equilibrium's axis-to-LCFS flux span. The three estimates use the reference density, prescribed temperature and common angular-velocity profiles.}
\label{tab:ion-label-maxima}
\begin{tabular}{@{}rrrrrrrr@{}}
\toprule
\(\Omega_0\) \((\mathrm{krad\,s^{-1}})\) & \(N_{\mathrm{He}}/N_e\) (\%) & \(k\) & Ion &
\(\lvert\delta x_s\rvert\) & \(\lvert\delta x_s N_{s,x}/N_s\rvert\) &
\(\lvert\delta x_s T_{s,x}/T_s\rvert\) & \(\lvert\delta x_s \Omega_{i,x}/\Omega_i\rvert\) \\
\midrule
70 & 0 & 1.00 & D & 0.50 & 0.23 & 1.14 & 1.19 \\
 &  &  & T & 0.75 & 0.34 & 1.71 & 1.79 \\
\addlinespace[1.5pt]
240 & 0 & 1.00 & D & 1.96 & 0.89 & 4.49 & 4.68 \\
 &  &  & T & 2.94 & 1.34 & 6.73 & 7.00 \\
\addlinespace[1.5pt]
246 & 0 & 1.00 & D & 2.05 & 0.93 & 4.68 & 4.87 \\
 &  &  & T & 3.06 & 1.39 & 7.01 & 7.29 \\
\addlinespace[1.5pt]
252 & 0 & 0.50 & D & 2.19 & 0.99 & 4.97 & 5.20 \\
 &  &  & T & 3.29 & 1.48 & 7.45 & 7.78 \\
\addlinespace[1.5pt]
252 & 0 & 1.00 & D & 2.16 & 0.98 & 4.95 & 5.12 \\
 &  &  & T & 3.24 & 1.46 & 7.41 & 7.68 \\
\addlinespace[1.5pt]
252 & 0 & 1.20 & D & 2.16 & 0.98 & 4.95 & 5.12 \\
 &  &  & T & 3.23 & 1.46 & 7.41 & 7.67 \\
\addlinespace[1.5pt]
252 & 2.5 & 1.00 & D & 2.14 & 0.97 & 4.90 & 5.08 \\
 &  &  & T & 3.21 & 1.45 & 7.33 & 7.61 \\
 &  &  & He & 2.13 & 0.96 & 4.86 & 5.05 \\
\addlinespace[1.5pt]
252 & 5 & 0.50 & D & 2.14 & 0.97 & 4.85 & 5.08 \\
 &  &  & T & 3.20 & 1.45 & 7.26 & 7.60 \\
 &  &  & He & 2.12 & 0.96 & 4.82 & 5.04 \\
\addlinespace[1.5pt]
252 & 5 & 1.00 & D & 2.12 & 0.96 & 4.86 & 5.05 \\
 &  &  & T & 3.18 & 1.44 & 7.27 & 7.56 \\
 &  &  & He & 2.11 & 0.96 & 4.83 & 5.02 \\
\addlinespace[1.5pt]
252 & 5 & 1.20 & D & 2.12 & 0.96 & 4.86 & 5.04 \\
 &  &  & T & 3.18 & 1.44 & 7.28 & 7.55 \\
 &  &  & He & 2.11 & 0.95 & 4.83 & 5.01 \\
\bottomrule
\end{tabular}
\end{table}

All four quantities in Table~\ref{tab:ion-label-maxima} attain their largest values for tritium in the pure-DT case at \(\Omega_0=252\ \mathrm{krad\,s^{-1}}\) and \(k=0.50\). The largest offset is \(3.29\%\) of the flux span. The three estimates peak at \(1.48\%\), \(7.45\%\) and \(7.78\%\) for the reference density, temperature and angular velocity, respectively. These values measure the sensitivity of the prescribed profiles to this offset on the saved grids.
\FloatBarrier

\section{Numerical acceptance criteria and residual checks}
\label{app:numerical-checks}

For the iterated and model current vectors \(\mathbf i\) and \(\widehat{\mathbf i}\) defined in Section~\ref{sec:2-3}, the source-consistency criterion is

\begin{equation}
r_J=\frac{\|\mathbf i-\widehat{\mathbf i}\|_2}
{\max\!\left(\|\widehat{\mathbf i}\|_2,\ |I_p^{\mathrm{ref}}|/\sqrt{N}\right)}
\leq 10^{-8},
\label{eq:acceptance-current-residual}
\end{equation}

where \(\|\cdot\|_2\) is the unweighted Euclidean norm. Calculations with prescribed total current or prescribed \(\gamma\) use the same normalisation for \(r_J\), with \(I_p^{\mathrm{ref}}\) as the reference current scale. In the fixed-current cases reported here, this scale equals the target current.

For both models, the unweighted \(L_2\) norm of the difference between the plasma-generated fluxes reconstructed from \(\mathbf i\) and \(\widehat{\mathbf i}\) must be at most \(10^{-7}\ \mathrm{Wb\,rad^{-1}}\) over the full grid.

The original discrete Grad-Shafranov operator acts on the plasma flux from \(\mathbf i\), and \(\widehat{\mathbf i}\) supplies the right-hand side. \(E_{\mathrm{GS}}\) is the ratio of the \(L_2\) norms of the residual and right-hand side at the interior grid points.

At the computational boundary, \(E_G\) is the \(L_2\) norm of the difference between that flux and the direct Green sum from \(\widehat{\mathbf i}\), normalised by the norm of the sum. Both residuals must be at most \(10^{-6}\).

The MF calculation additionally requires \(E_Q\le 2\times10^{-13}\), where \(E_Q\) is the maximum of \(\lvert\sum_s Z_s n_s\rvert/\sum_s |Z_s|n_s\) over the plasma evaluation points. For reference densities with a shared factor \(h(x)\) that vanishes at \(x=1\), the edge ratio is the interior limit after cancelling \(h\). For fixed-current cases in either model, the total currents from \(\mathbf i\) and \(\widehat{\mathbf i}\) must each lie within \(6\ \mathrm A\) of the target. When \(\gamma\) is prescribed, the converged total current is an output.

Table~\ref{tab:acceptance-residuals} reports the normalised original discrete Grad-Shafranov residual \(E_{\mathrm{GS}}\), Green-boundary residual \(E_G\), source-consistency residual \(E_J=r_J\), maximum relative quasineutrality residual \(E_Q\), and reconstructed-flux difference \(E_{\psi,\mathrm{rec}}\). The source-consistency residual is defined in Eq.~\eqref{eq:acceptance-current-residual}. The last column gives the larger absolute error of the total currents obtained from \(\mathbf i\) and \(\widehat{\mathbf i}\) relative to the prescribed \(I_p\).

\begin{table}[!tbp]
\centering
\small
\setlength{\tabcolsep}{3pt}
\renewcommand{\arraystretch}{1.12}
\caption{Equilibrium-equation residuals and current errors for the grid-refinement calculations of Section~\ref{sec:3-3}.}
\label{tab:acceptance-residuals}
\fittable{\begin{tabular}{@{}llrrrrrr@{}}
\toprule
Plasma & Grid & $E_{GS}$ & $E_G$ & $E_J$ & $E_Q$ & \shortstack{$E_{\psi,rec}$\\Wb rad\textsuperscript{-1}} & \shortstack{$\lvert\delta I_p\rvert$\\A} \\
\midrule
DT & \(129\times257\) & \(1.64\times10^{-10}\) & \(2.19\times10^{-11}\) & \(1.57\times10^{-10}\) & \(7.11\times10^{-15}\) & \(5.76\times10^{-8}\) & \(2.82\times10^{-4}\) \\
DT & \(193\times385\) & \(8.87\times10^{-12}\) & \(5.09\times10^{-14}\) & \(1.46\times10^{-12}\) & \(1.78\times10^{-14}\) & \(2.75\times10^{-10}\) & \(1.94\times10^{-7}\) \\
DT & \(257\times513\) & \(1.57\times10^{-11}\) & \(4.59\times10^{-13}\) & \(5.81\times10^{-12}\) & \(7.82\times10^{-14}\) & \(2.43\times10^{-9}\) & \(2.31\times10^{-6}\) \\
DT--He & \(129\times257\) & \(7.14\times10^{-12}\) & \(7.44\times10^{-13}\) & \(6.00\times10^{-12}\) & \(9.95\times10^{-14}\) & \(1.67\times10^{-9}\) & \(9.28\times10^{-6}\) \\
DT--He & \(193\times385\) & \(8.51\times10^{-12}\) & \(7.91\times10^{-14}\) & \(1.24\times10^{-12}\) & \(9.95\times10^{-14}\) & \(3.56\times10^{-10}\) & \(8.01\times10^{-7}\) \\
DT--He & \(257\times513\) & \(4.82\times10^{-11}\) & \(2.96\times10^{-12}\) & \(4.70\times10^{-11}\) & \(9.95\times10^{-14}\) & \(1.94\times10^{-8}\) & \(2.64\times10^{-5}\) \\
\bottomrule
\end{tabular}}
\par\smallskip{\footnotesize\raggedright Each entry is the maximum over the two models at the stated grid, except \(E_Q\), which applies to the multi-fluid model. The reconstructed-flux measure compares the flux distributions reconstructed from \(\mathbf i\) and \(\widehat{\mathbf i}\). The current error includes both integrated currents.\par}
\end{table}

Across the 10 limiting-case calculations in Section~\ref{sec:3-2} and the 12 grid-refinement calculations in Section~\ref{sec:3-3}, the maximum original Grad-Shafranov, Green-boundary and source-consistency residuals are \(1.19\times10^{-9}\), \(4.40\times10^{-10}\) and \(1.14\times10^{-9}\), respectively. The maximum quasineutrality residual in the MF calculations is \(9.95\times10^{-14}\). The maximum reconstructed-flux difference is \(5.76\times10^{-8}\ \mathrm{Wb\,rad^{-1}}\), and the maximum absolute plasma-current error is \(2.82\times10^{-4}\ \mathrm A\).
\FloatBarrier

\section[Analytical relations between multi-fluid and rotating MHD pressures]{Analytical relations between multi-fluid\\and rotating MHD pressures}
\label{app:pressure-ratio}

\subsection*{General pressure inequality and equality condition}

We evaluate the pressure expressions in Eqs.~\eqref{eq:2-3} and \eqref{eq:2-7} at the same \(R\), \(x\) and reference profiles, suppressing the \(x\) dependence in the notation. All ions share \(\Omega\). As in Section~\ref{sec:2-1}, \(Z_e=-1\) and neglect of electron inertia gives \(W_e=0\). Take \(N_s\geq0\), \(\theta_s>0\), \(P_0>0\) and \(\sum_s Z_s N_s=0\). The species sums below include electrons. From Eqs.~\eqref{eq:2-2} and \eqref{eq:2-6}, define the reference pressure fractions and density exponents
\[
w_s=\frac{N_s\theta_s}{P_0},\qquad
g_s=\frac{W_s-Z_s e\Phi}{\theta_s},\qquad
\sum_s w_s=1.
\]
With \(n_s=N_s e^{g_s}\), convexity of the exponential gives
\[
\frac{P_{\mathrm{MF}}}{P_0}
=\sum_s w_s e^{g_s}
\geq \exp\!\left(\sum_s w_s g_s\right).
\]
Reference quasineutrality cancels the electrostatic contribution to the weighted mean. Since \(W_e=0\), the centrifugal contribution uses the ion mass density \(\rho_0\).
\[
\sum_s w_s g_s
=\frac{\sum_s N_s W_s-e\Phi\sum_s Z_s N_s}{P_0}
=\frac{\rho_0\Omega^2(R^2-R_0^2)}{2P_0}.
\]
This is the exponent in Eq.~\eqref{eq:2-7}. It follows that
\begin{equation}
\boxed{P_{\mathrm{MF}}(R,x)\geq P_{\mathrm{MHD}}(R,x)}.
\label{eq:app-pressure-bound}
\end{equation}

By strict convexity, equality in Eq.~\eqref{eq:app-pressure-bound} holds exactly when all species with \(N_s>0\) have \(g_s=c\) for a common \(c\), or equivalently share \(n_s/N_s\). The electron exponent gives \(e\Phi=c\theta_e\), and each ion satisfies \(\tfrac12m_s\Omega^2(R^2-R_0^2)=c(\theta_s+Z_s\theta_e)\). For \(\Omega^2(R^2-R_0^2)\ne0\), the two pressure expressions coincide exactly when there is a value \(a\) common to all ions present such that
\begin{equation}
\boxed{\frac{m_s}{\theta_s+Z_s\theta_e}=a},
\qquad s\ne e,\quad N_s>0.
\label{eq:app-pressure-equality}
\end{equation}
Conversely, Eq.~\eqref{eq:app-pressure-equality} gives \(c=\tfrac12 a\Omega^2(R^2-R_0^2)\) and hence \(n_s=N_s e^c\). The common factor preserves quasineutrality. The charge sum in Eq.~\eqref{eq:2-3} decreases strictly with \(\Phi\), fixing the solution. A single ion species meets the condition automatically. The mass difference between singly charged D and T at a common temperature makes the inequality strict for \(\Omega\ne0\) and \(R\ne R_0\). When \(\Omega^2(R^2-R_0^2)=0\), quasineutrality gives \(\Phi=0\), so both pressures equal \(P_0\).

For finite rotation, fix \(R\), \(x\) and the reference densities and temperatures. Define \(\zeta=\tfrac12\Omega^2(R^2-R_0^2)\) and \(u=e\Phi\). Before imposing quasineutrality, the total pressure is
\[
\mathcal P(\zeta,u)=N_e\theta_e e^{u/\theta_e}
+\sum_i N_i\theta_i e^{(m_i\zeta-Z_i u)/\theta_i},
\]
where \(i\) runs over the ion species and \(\theta_i\) is the thermal energy of species \(i\). Quasineutrality is the stationary condition in \(u\), with positive curvature.
\[
\partial_u\mathcal P=n_e-\sum_i Z_i n_i=0,
\qquad
\partial_u^2\mathcal P=\frac{n_e}{\theta_e}+\sum_i\frac{Z_i^2 n_i}{\theta_i}>0.
\]
At fixed \(\zeta\), electrons and positive ions give \(\mathcal P\to\infty\) as \(|u|\to\infty\). Hence the quasineutral solution \(u\) is the unique minimiser of \(\mathcal P\).

The function \(\mathcal F(\zeta,u)=\ln[\mathcal P(\zeta,u)/P_0]\) is a log-sum-exp of affine functions of \((\zeta,u)\) for the species with positive reference densities, so it is jointly convex. Partial minimisation over \(u\) preserves convexity in \(\zeta\), so
\[
f(\zeta)=\min_u\mathcal F(\zeta,u)=\ln\frac{P_{\mathrm{MF}}(\zeta)}{P_0}
\]
is convex. At \(\zeta=0\), reference quasineutrality gives \(u=0\). Using stationarity in \(u\) to differentiate \(f\), and Eq.~\eqref{eq:2-7} for the rotating-MHD pressure, gives
\[
f(0)=0,\qquad f'(0)=\frac{\partial_\zeta\mathcal P(0,0)}{P_0}=\frac{\rho_0}{P_0},
\qquad
\ln\frac{P_{\mathrm{MHD}}}{P_0}=\frac{\rho_0}{P_0}\zeta.
\]
Hence
\[
L(\zeta)=\ln\frac{P_{\mathrm{MF}}}{P_{\mathrm{MHD}}}
=f(\zeta)-\frac{\rho_0}{P_0}\zeta
\]
is convex with \(L(0)=L'(0)=0\). Its value is nondecreasing as \(\zeta\) increases from zero on the positive half-axis or decreases from zero on the negative half-axis. Increasing \(|\Omega|\) follows these directions for \(R>R_0\) and \(R<R_0\), respectively, so the pressure ratio is nondecreasing in \(|\Omega|\) at the fixed coordinates and reference inputs. At \(R=R_0\), the ratio is identically one.

At fixed \(R\), \(x\) and reference profiles, scale \(\Omega\) by \(\lambda\). Write \(W_s=\lambda^2\widehat W_s\), with \(\widehat W_e=0\), and \(e\Phi=\lambda^2U_2+O(\lambda^4)\). Linearising quasineutrality gives \(U_2\), and \(h_s\) is the coefficient of \(\lambda^2\) in \(g_s\).
\[
U_2=\frac{\sum_s N_s Z_s\widehat W_s/\theta_s}{\sum_s N_s Z_s^2/\theta_s},
\qquad
h_s=\frac{\widehat W_s-Z_s U_2}{\theta_s}.
\]
Expanding the pressure ratio gives
\[
\frac{P_{\mathrm{MF}}}{P_{\mathrm{MHD}}}
=1+\frac{\lambda^4}{2}
\left[\sum_s w_s h_s^2-\left(\sum_s w_s h_s\right)^2\right]
+O(\lambda^6).
\]
The pressure-weighted variance of the \(h_s\) sets the leading fractional pressure difference.

\subsection*{Exact pressure ratio for pure DT with equal reference densities}

For pure DT with \(N_D=N_T=N_e/2\), both ions share thermal energy \(\theta_i\) and angular velocity \(\Omega\). With the same comparison conditions as above, define the factor in the ion density sum from Eq.~\eqref{eq:2-2}
\[
A=\tfrac12\left(e^{W_D/\theta_i}+e^{W_T/\theta_i}\right),\qquad
W_s=\tfrac12 m_s\Omega^2(R^2-R_0^2),\quad s\in\{D,T\}.
\]
The quasineutrality condition \(n_D+n_T=n_e\) in Eq.~\eqref{eq:2-3} then gives
\[
A e^{-e\Phi/\theta_i}=e^{e\Phi/\theta_e},\qquad
e\Phi=\frac{\theta_i\theta_e}{\theta_i+\theta_e}\ln A.
\]
Substitution into the electron density relation and the pressure sum gives
\[
n_e=N_e A^{\theta_i/(\theta_i+\theta_e)},\qquad
P_{\mathrm{MF}}=(\theta_i+\theta_e)n_e
=P_0 A^{\theta_i/(\theta_i+\theta_e)}.
\]
To separate the mean centrifugal energy and the D--T difference, write
\[
\overline W=\frac{W_D+W_T}{2},\qquad
\chi=\frac{W_T-W_D}{2\theta_i}.
\]
The sum of exponentials and the multi-fluid pressure become
\[
A=e^{\overline W/\theta_i}\cosh\chi,\qquad
P_{\mathrm{MF}}=P_0 e^{\overline W/(\theta_i+\theta_e)}
[\cosh\chi]^{\theta_i/(\theta_i+\theta_e)}.
\]
The same reference profiles give
\[
P_0=N_e(\theta_i+\theta_e),\qquad
\rho_0=\tfrac12 N_e(m_D+m_T).
\]
Using these profiles in the rotating-MHD pressure in Eq.~\eqref{eq:2-7} yields
\begin{equation}
P_{\mathrm{MHD}}=P_0 e^{\overline W/(\theta_i+\theta_e)},\qquad
\frac{P_{\mathrm{MF}}}{P_{\mathrm{MHD}}}
=[\cosh\chi]^{\theta_i/(\theta_i+\theta_e)}.
\label{eq:app-dt-pressure-ratio}
\end{equation}

Expanding Eq.~\eqref{eq:app-dt-pressure-ratio} at \(\chi=0\) and taking the ratio of ion densities from Eq.~\eqref{eq:2-2} give
\[
\frac{P_{\mathrm{MF}}}{P_{\mathrm{MHD}}}-1
=\frac{\theta_i}{2(\theta_i+\theta_e)}\chi^2+O(\chi^4),
\qquad
\frac{n_T}{n_D}=e^{2\chi},
\qquad
\ln S_{T/D}=2(\chi_{\mathrm{out}}-\chi_{\mathrm{in}}).
\]
The local D and T densities differ wherever \(\chi\ne0\). Scaling \(\Omega\) by \(\lambda\) at fixed geometry and reference profiles gives \(\chi=O(\lambda^2)\). The leading terms in D--T separation and in the pressure difference are therefore of orders \(\lambda^2\) and \(\lambda^4\), respectively.
\FloatBarrier

\clearpage

\begingroup
\linespread{1}\selectfont
\fontsize{10}{12}\selectfont
\setlength{\bibsep}{0.75pt plus 0.5pt minus 0.5pt}
\bibliographystyle{unsrtnat}
\bibliography{references}
\endgroup
\end{document}